\documentclass[12pt]{article}
\usepackage{xcolor}
\usepackage{amsmath,amsthm}
\usepackage{graphicx,psfrag,epsf}
\usepackage{enumerate}
\usepackage{natbib}
\usepackage{url} 
\usepackage{enumitem} 
\usepackage{booktabs} 
\usepackage{multirow} 
\usepackage{array} 
\usepackage{tikz}
\usetikzlibrary{matrix, arrows.meta, positioning}
\usetikzlibrary{calc}
\usetikzlibrary{shapes.geometric, patterns, fit}
\usepackage{pgfmath}
\usepackage{caption}
\usepackage{subcaption}
\usepackage{amsmath,color}
\usepackage{float}
\usepackage{algorithm}
\usepackage{algorithmic}
\usepackage{amsfonts}
\usepackage{appendix}
\usepackage{bbm}
\usepackage{amssymb}
\usepackage{mathabx}
\usepackage[colorlinks=true, linkcolor=blue, citecolor=blue, urlcolor=blue]{hyperref}
\allowdisplaybreaks

\newtheorem{remark}{Remark}
\newtheorem{theorem}{Theorem}

\newtheorem{definition}{Definition}

\newcommand{\blind}{1}

\usepackage[top=1in, bottom=1in, left=1in, right=1in]{geometry}

\begin{document}

\def\spacingset#1{\renewcommand{\baselinestretch}%
{#1}\small\normalsize} \spacingset{1}
\def\red{\color{red}}
\def\olive{} % color for update
\def\orange{\color{orange}} 

%%%%%%%%%%%%%%%%%%%%%%%%%%%%%%%%%%%%%%%%%%%%%%%%%%%%%%%%%%%%%%%%%%%%%%%%%%%%%%

\if1\blind
{
  \title{\bf Two-sample Testing with Block-wise Missingness in Multi-source Data}
  \author{
    Kejian Zhang$^{1}$, 
    Muxuan Liang$^{2}$, 
    Robert Maile$^{3}$, 
    Doudou Zhou$^{1,\dagger}$ \\
    \\
{\footnotesize    {$^{1}$Department of Statistics and Data Science, National University of Singapore, Singapore}} \\
   {\footnotesize {$^{2}$Department of Biostatistics, The University of Texas MD Anderson Cancer Center, USA} }\\
{\footnotesize    {$^{3}$Department of Surgery, College of Medicine, University of Florida, USA}} \\
  {\footnotesize  {$^{\dagger}$Corresponding author:  ddzhou@nus.edu.sg}}
  }
  \date{}
  \maketitle
} \fi

\if0\blind
{
  \bigskip
  \bigskip
  \bigskip
  \begin{center}
    {\LARGE\bf Two-sample Testing with Block-wise Missingness in Multi-source Data}
\end{center}
  \medskip
} \fi

\begin{abstract}
Multi-source and multi-modal datasets are increasingly common in scientific research, yet they often exhibit  \emph{block-wise missingness}, where entire modalities are systematically absent in some sources or no single source contains all modalities. This structured missingness poses major challenges for two-sample hypothesis testing. Standard approaches, such as imputation or complete-case analysis, may introduce bias or suffer efficiency loss, especially under missingness-not-at-random mechanisms. To address this challenge, we propose the Block-Pattern Enhanced Test, a general framework for constructing two-sample testing statistics that explicitly accounts for block-wise missingness. We show that the framework yields valid tests under a new condition allowing for missing-not-at-random mechanism. Building on this general framework, we further propose the Block-wise Rank In Similarity graph Edge-count (BRISE) test, which accommodate heterogeneous modalities using rank-based similarity graphs. Theoretically, we establish that the null distribution of BRISE converges to a $\chi^2$ distribution, and that the test is consistent both in the standard asymptotic regime and in the high-dimensional low-sample-size setting under mild conditions. Simulation studies demonstrate that BRISE controls the type-I error rate and achieves strong power across a wide range of alternatives. Applications to two real-world datasets with block-wise missingness further illustrate the practical utility of the proposed method.
\end{abstract}

\noindent%
\vfill

\spacingset{1.45} % DON'T change the spacing!

\section{Introduction}
\label{section:intro}

Multi-source and multi-modal datasets are increasingly common in scientific research, encompassing diverse data types collected across individuals, institutions, and time. For example, the Alzheimer’s Disease Neuroimaging Initiative (ADNI) study \citep{ADNIa,ADNIb} combines multiple modalities of various types from multiple sources, including imaging, genetic, clinical data, and electronic health records (EHRs) \citep{MIMIC-IV} that integrate medication histories and clinical notes to understand possible causes and progressions of Alzheimer’s Disease in different patient groups. In such scientific research, a fundamental task in statistics is to determine whether two pre-specified groups/samples come from the same distribution. 
Formally, let  $X_1,\ldots,X_m\overset{i.i.d}{\sim}F_X$ and $Y_1,\ldots, Y_n\overset{i.i.d}{\sim}F_Y$ be two independent samples. The goal is to test
\[
H_0: F_X = F_Y \quad \text{versus} \quad H_a: F_X \neq F_Y.
\]
In this work, we focus on two-sample problem for multi-source and multi-modal data. 

\begin{figure}[h]
    \centering
    \begin{tikzpicture}[scale=0.8, every node/.style={font=\small}] 
    % Define colors for the cells
    \definecolor{red}{RGB}{217, 83, 25}      
    \definecolor{orange}{RGB}{237, 177, 32} 
    \definecolor{yellow}{RGB}{254, 217, 118}  
    \definecolor{lyellow}{RGB}{255, 255, 204}  
    \definecolor{blue}{RGB}{8, 81, 156}    
    \definecolor{green}{RGB}{119, 172, 48}    
    \definecolor{cyan}{RGB}{107, 174, 214}  

    % Draw the main table (7x3 grid)
    \foreach \row/\colorA/\colorB/\colorC in {
        7/red/red/red,
        6/orange/orange/gray!20,
        5/yellow/gray!20/yellow,
        4/gray!20/lyellow/lyellow,
        3/blue/gray!20/gray!20,
        2/gray!20/green/gray!20,
        1/gray!20/gray!20/cyan} {
        \fill[\colorA] (0,\row) rectangle (7/3,\row-1); % First column
        \fill[\colorB] (7/3,\row) rectangle (14/3,\row-1); % Second column
        \fill[\colorC] (14/3,\row) rectangle (7,\row-1); % Third column
    }

    % Highlight the blocks
    \draw[ultra thick, dashed] (7/3,2) rectangle (14/3,1);
    \draw[ultra thick, dashed] (7,6) rectangle (14/3,5);

    % Draw the grid lines
    \draw[black] (0,7) rectangle (7,0);
    \foreach \x in {7/3,14/3} \draw[black] (\x,0) -- (\x,7);
    \foreach \y in {1,2,3,4,5,6} \draw[black] (0,\y) -- (7,\y);

    % Add group and source labels
    \node[anchor=south] at (7/6,7.2) {\textit{Source~1}};
    \node[anchor=south] at (21/6,7.2) {\textit{Source~2}};
    \node[anchor=south] at (35/6,7.2) {\textit{Source~3}};

    \node[anchor=east] at (-0.2,6.5) {Pattern 1 ($\mathcal{P}_1$)};
    \node[anchor=east] at (-0.2,5.5) {Pattern 2 ($\mathcal{P}_2$)};
    \node[anchor=east] at (-0.2,4.5) {Pattern 3 ($\mathcal{P}_3$)};
    \node[anchor=east] at (-0.2,3.5) {Pattern 4 ($\mathcal{P}_4$)};
    \node[anchor=east] at (-0.2,2.5) {Pattern 5 ($\mathcal{P}_5$)};
    \node[anchor=east] at (-0.2,1.5) {Pattern 6 ($\mathcal{P}_6$)};
    \node[anchor=east] at (-0.2,0.5) {Pattern 7 ($\mathcal{P}_7$)};

    % Add a callout to the highlighted block
    \draw[-Stealth, ultra thick] (14/3,1.5) -- (8,1.5);
    \draw[-Stealth, ultra thick] (7,5.5) -- (8,5.5);

    % Fill cells to represent data
    \fill[green!30] (9,0.5) rectangle (12,2.5);
    \fill[gray!10] (9,4.5) rectangle (12,6.5);
    
    % Add details about the highlighted block on the right
    \draw (9,2.5) rectangle (12,0.5);
    \foreach \x in {9.5,10,10.5,11,11.5} \draw (\x,0.5) -- (\x,2.5); % Vertical lines
    \foreach \y in {1,1.5,2} \draw (9,\y) -- (12,\y); % Horizontal lines
    \draw (9,4.5) rectangle (12,6.5);
    \foreach \x in {9.5,10,10.5,11,11.5} \draw (\x,4.5) -- (\x,6.5); % Vertical lines
    \foreach \y in {5,5.5,6} \draw (9,\y) -- (12,\y); % Horizontal lines

    % Labels for observations and dimensions
    \draw[->, thick] (8.8,2.7) -- ++(0,-2.5) node[midway, rotate=90, above=4pt] {Observation};
    \draw[->, thick] (8.8,2.7) -- ++(3.5,0) node[midway, above=4pt] {Dimension};
    \draw[->, thick] (8.8,6.7) -- ++(0,-2.5) node[midway, rotate=90, above=4pt] {Observation};
    \draw[->, thick] (8.8,6.7) -- ++(3.5,0) node[midway, above=4pt] {Dimension};

    \end{tikzpicture}
    \caption{Pattern partition across three sources, with gray blocks indicating missing data. Seven nonempty patterns are shown: Pattern 1 is fully observed; Patterns 2–4 each have one missing source; Patterns 5–7 each have two missing sources.}
    \label{fig:example_3sources}
\end{figure}
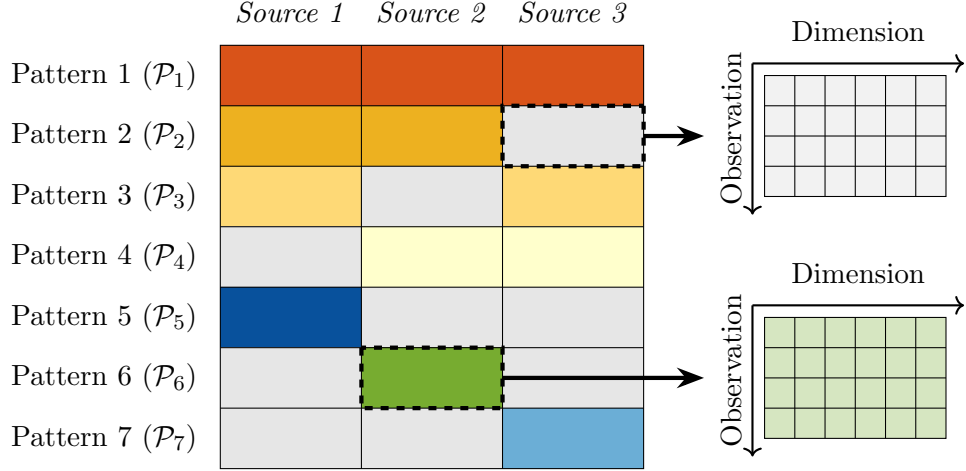

With more scientific research integrate multiple modalities from multiple sources, unsatisfactory data quality and structure are common and may bring challenges in statistical analysis, especially for two-sample testing.
A key issue is \emph{block-wise missingness}, where entire data modalities are absent for certain sources (Figure~\ref{fig:example_3sources}); and it is possible that no single source has all modalities available. This data structure often stems from non-uniform clinical priorities, resource constraints, or institutional protocols; and thus, the missing may not be at random but associated with the study itself. For instance, in ADNI study, site-level practices can produce systematic gaps across modalities \citep{F.Xue2020}. Similar missing patterns appear in partial or incomplete multi-view learning \citep{li2014partial,zhang2020deep,zhou2023multi}, such as bilingual document clustering with missing translations or audiovisual tasks where only one modality is available.

\begin{figure}[t]
% Define colors for the cells
\definecolor{red}{RGB}{217, 83, 25}      
\definecolor{orange}{RGB}{237, 177, 32} 
\definecolor{yellow}{RGB}{254, 217, 118}  
\definecolor{lyellow}{RGB}{255, 255, 204}  
\definecolor{blue}{RGB}{8, 81, 156}    
\definecolor{green}{RGB}{119, 172, 48}    
\definecolor{cyan}{RGB}{107, 174, 214}   
\centering
\tikzset{
    cell/.style={rectangle, draw=black, minimum width=0.4cm, minimum height=0.4cm},
    filled/.style={cell, fill=#1},
    missing/.style={cell, fill=gray!20},
    imputed/.style={cell, fill=purple!20, draw=blue, dashed},
    arrowstyle/.style={-Stealth, ultra thick},
}

\begin{tikzpicture}[node distance=2cm and 2cm]

%%%%% Original Matrix %%%%%
\matrix (original) [matrix of nodes, nodes={cell}, column sep=0pt, row sep=0pt] {
    \node[filled=red] {}; & \node[filled=red] {}; & \node[filled=red] {}; & \node[filled=red] {}; & \node[filled=red] {}; \\
    \node[filled=red] {}; & \node[filled=red] {}; & \node[filled=red] {}; & \node[filled=red] {}; & \node[filled=red] {}; \\
    \node[filled=orange] {}; & \node[filled=orange] {}; & \node[filled=orange] {}; & \node[missing] {}; & \node[filled=orange] {}; \\
    \node[filled=blue] {}; & \node[filled=blue] {}; & \node[missing] {}; & \node[missing] {}; & \node[filled=blue] {}; \\
    \node[filled=green] {}; & \node[filled=green] {}; & \node[filled=green] {}; & \node[filled=green] {}; & \node[missing] {}; \\
};
% Axes for Original Data
\draw[->, thick] (original.north west) -- ++(0,-2.5) node[midway, rotate=90, above=4pt] {Observation};
\draw[->, thick] (original.north west) -- ++(2.5,0) node[midway, above=4pt] {Dimension};

\node[above=0.8cm] at (original.north) {\textbf{Original Data}};

\coordinate (origAnchor) at (original.east);

%%%%% Complete-case Matrix %%%%%
\matrix (case) [matrix of nodes, nodes={cell}, column sep=0pt, row sep=0pt, right=2cm of original.east, anchor= west] {
    \node[filled=red] {}; & \node[filled=red] {}; & \node[filled=red] {}; & \node[filled=red] {}; & \node[filled=red] {}; \\
    \node[filled=red] {}; & \node[filled=red] {}; & \node[filled=red] {}; & \node[filled=red] {}; & \node[filled=red] {}; \\
};
\node[right=0.4cm] at (case.east) {(b) Complete-case};
\coordinate (caseAnchor) at (case.west);

%%%%% Imputation Matrix %%%%%
\matrix (impute)  [matrix of nodes, nodes={cell}, column sep=0pt, row sep=0pt, above=0.5cm of case]{
    \node[filled=red] {}; & \node[filled=red] {}; & \node[filled=red] {}; & \node[filled=red] {}; & \node[filled=red] {}; \\
    \node[filled=red] {}; & \node[filled=red] {}; & \node[filled=red] {}; & \node[filled=red] {}; & \node[filled=red] {}; \\
    \node[filled=orange] {}; & \node[filled=orange] {}; & \node[filled=orange] {}; & \node[imputed] {}; & \node[filled=orange] {}; \\
    \node[filled=blue] {}; & \node[filled=blue] {}; & \node[imputed] {}; & \node[imputed] {}; & \node[filled=blue] {}; \\
    \node[filled=green] {}; & \node[filled=green] {}; & \node[filled=green] {}; & \node[filled=green] {}; & \node[imputed] {}; \\
};
\node[right=0.4cm] at (impute.east) {(a) Imputation};
\coordinate (impAnchor) at (impute.west);

%%%%% Complete-dimension Matrix %%%%%
\matrix (dim) [matrix of nodes, nodes={cell}, column sep=0pt, row sep=0pt, below=0.5cm of case] {
    \node[filled=red] {}; & \node[filled=red] {}; \\
    \node[filled=red] {}; & \node[filled=red] {}; \\
    \node[filled=orange] {}; & \node[filled=orange] {}; \\
    \node[filled=blue] {}; & \node[filled=blue] {}; \\
    \node[filled=green] {}; & \node[filled=green] {}; \\
};
\node[right=1cm] at (dim.east) {(c) Available-case};
\coordinate (dimAnchor) at (dim.west);

%%%%% Arrows — adjustable %%%%%
\draw[arrowstyle] (origAnchor) -- (impAnchor);
\draw[arrowstyle] (origAnchor) -- (caseAnchor);
\draw[arrowstyle] (origAnchor) -- (dimAnchor);

\end{tikzpicture}

\caption{Three common strategies for handling missing data (gray entries).
}
\label{fig:Methods for handling missing data}
\end{figure}
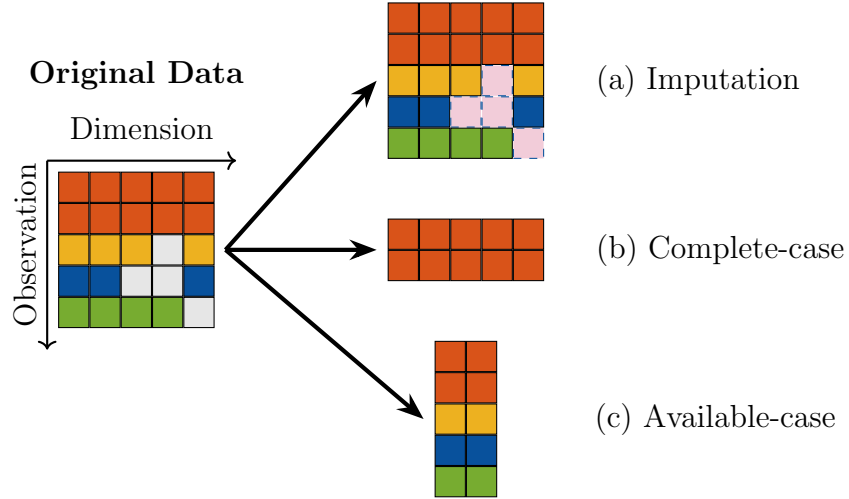

This challenge is especially acute in the context of two-sample testing. However, nonparametric tests for missing data are scarce; naive implementation of existing methods leads to inflated type-I error or unsatisfied power. Figure~\ref{fig:Methods for handling missing data} summarizes three common existing strategies. Specifically, imputation method fills in missing entries using model-based predictions \citep{Rubin1990multiImputation,Zhang2016SingleImputation} but relies on strong assumptions, which can bias inference if those assumptions fail. Complete-case analysis discards incomplete samples, which reduces effective sample size \citep{Little1992Completecase,Ross2020Completecase}. Available-case analysis uses only variables required for a given comparison but can still suffer information loss and inefficiency \citep{Little1992Completecase}. Recent work has extended classical two-sample testing procedures to handle entrywise missingness, but these approaches often require missing-at-random (MAR) assumption and are conservative under block-wise missingness \citep{zeng2024mmd, Y.Zeng2024, MMD, aleksic2025twosampletestingmissingdata}. Thus, a method tailored to structured block-wise mechanisms, possibly accommodating missing-not-at-random (MNAR), is still needed.

Another challenge comes from integrating multiple modalities of heterogeneous data types. In the example of the ADNI study, multiple modalities are combined,  including both Euclidean and non-Euclidean data. Structured data, such as clinical variables and genetic information, typically lie in Euclidean domains, whereas unstructure data, such as image data and clinical notes, may not share the same geometry as Euclidean data. As a result, a testing procedure should incorporate general and flexible distance metrics, rather than relying solely on Euclidean distances, to effectively integrate modalities of different data types in statistical analysis.

When the data are fully observed, many powerful nonparametric methods are available, including graph-based tests \citep{1979.Friedman,schilling1986multivariate,Henze1988,rosenbaum2005exact,chen2013graph,chen2017new,chen2018weighted, D.Zhou2023}, classification-based tests \citep{lopez2016revisiting,kim2021classification,HEDIGER2022}, inter-point distance-based tests \citep{biswas2014nonparametric,li2018asymptotic}, and kernel-based tests \citep{NIPS2006_e9fb2eda,eric2007testing,gretton2009fast,MMD,song2024generalized}. Therefore, a method capable of integrating both Euclidean and non-Euclidean modalities has the potential to substantially enhance the utility of multi-source and multi-modal data.

In this work, to address structured block-wise missing mechanisms and accommodate possible MNAR missingness, we propose the \textbf{B}lock-\textbf{P}attern \textbf{E}nhanced \textbf{T}est (BPET), a general framework for two-sample testing under block-wise missing data. BPET partitions the dataset according to missingness patterns, applies pattern-aware procedures to pairs of patterns using shared observed sources, and aggregates the results into a global test statistic, allowing direct inference on partially observed data without imputation or case deletion. Importantly, BPET is not tied to any specific testing method, enabling a broad class of two-sample tests to be adapted within the framework. We show that BPET yields valid testing statistics under assumptions strictly weaker than Missing Completely At Random (MCAR), while allowing for MNAR missingness.

\newpage 

To better integrate modalities of heterogeneous types, we further apply the proposed framework with a rank-weighted similarity graph-based testing method \citep{D.Zhou2023} and develop a specific test, termed the \textbf{B}lock-wise \textbf{R}ank \textbf{I}n \textbf{S}imilarity graph \textbf{E}dge-count (BRISE) test. Under mild regularity conditions, we establish that the BRISE statistic converges to a $\chi^2$ distribution under the null hypothesis and that the test is consistent in both the standard asymptotic regime and the high-dimensional low-sample-size setting.  

For finite-sample inference, we develop a tailored permutation-based procedure to approximate the null distribution. Unlike standard permutation, which shuffles all labels while fixing only total group sizes, we introduce a \emph{pattern-wise permutation} that restricts permutations within each missingness pattern. We show that standard permutation can fail when groups have unequal marginal pattern distributions, whereas the proposed pattern-wise permutation yields valid inference.

In summary, our contributions are threefold: (1)  We propose BPET, a general framework for two-sample testing on partially observed multi-source and multi-modal data that operates directly on block-wise missing data without imputation or case deletion; (2) Within the proposed framework, we develop the BRISE test by integrating rank-weighted similarity graph methods to accommodate modalities of heterogeneous data types; and (3) We establish the asymptotic $\chi^2$ null distribution and consistency of BRISE in both standard asymptotic and high-dimensional low-sample-size regimes, and propose a pattern-wise permutation scheme for valid finite-sample inference when standard permutation may fail.

The remainder of the paper is organized as follows. Section~\ref{section:method} presents BPET and Section~\ref{subsec:NNG-induced Rank} introduces BRISE; Section~\ref{section: asymptotic properties} develops theory; Sections~\ref{section: simluations} and~\ref{section: real data} present numerical and real data results; and Section~\ref{section: discussion} concludes.

\section{Framework of Block-Pattern Enhanced Test}
\label{section:method}

In this section, we introduce the general framework of BPET and a pattern-wise permutation procedure to approximate the null distribution of the test statistic. We also establish the validity of the framework under different missingness assumptions, including possible MNAR mechanisms.

\subsection{General Framework}
\label{subsec: BPET wf}
The key idea of BPET is to construct pattern-aware test statistics and aggregate them into a global statistic for two-sample testing. Since each pattern-specific statistic involves only distributions conditional on a given missingness pattern, BPET yields a valid global test as long as, under the null hypothesis, the two samples share the same conditional distribution within each pattern, allowing for possible MNAR missingness.

\paragraph{Stage 1: Missing-Pattern Partition.}
Suppose there are $L$ sources or modalities, indexed by $l = 1, \dots, L$, where each modality is defined on a domain $\mathcal{S}^{(l)}$. These domains may vary across modalities, ranging from Euclidean spaces such as $\mathbb{R}^{d_l}$ to non-Euclidean spaces such as networks or functional data spaces. Let $n_P$ denote the number of distinct missingness patterns observed in the data. Each missing pattern is represented as 
\[
\mathcal{P}_\alpha = (P_\alpha^{(1)}, \ldots, P_\alpha^{(L)}), \quad \alpha \in \{1, \dots, n_P\},
\]
where $P_\alpha^{(l)} = 1$ if modality $l$ is observed and 0 otherwise. For instance, when $L = 3$, Figure~\ref{fig:example_3sources} shows that $\mathcal{P}_1 = (1,1,1)$ corresponds to complete observation; $\mathcal{P}_2=(1,1,0)$, $\mathcal{P}_3=(1,0,1)$, $\mathcal{P}_4=(0,1,1)$ correspond to cases with two observed modalities;  and $\mathcal{P}_5=(1,0,0)$, $\mathcal{P}_6=(0,1,0)$, $\mathcal{P}_7=(0,0,1)$ correspond to cases with only one observed modality. In general, the number of possible nonempty patterns is at most $2^L - 1$.

To unify notation, let $\{Z_i\}_{i=1}^{N}$ denote the pooled sample, where $\{Z_i\}_{i=1}^{m} = \{X_j\}_{j=1}^m$, $\{Z_i\}_{i=m+1}^{N} = \{Y_j\}_{j=1}^{n}$, and $N = m + n$. Throughout, we assume block-wise missingness: for each observation, variables within a modality are either fully observed or fully missing. Thus, each observation can be written as $Z_i = (Z_i^{(1)}, \ldots, Z_i^{(L)})$, where $Z_i^{(l)} \in \mathcal{S}^{(l)}$ if observed, and its missingness pattern is denoted by $S_i = (S_i^{(1)}, \ldots, S_i^{(L)}) \in \{0,1\}^L$. 

The set of observations following pattern $\mathcal{P}_\alpha$ is
\[
\mathcal{Z}^{(\alpha)} = \{ Z_i : S_i = \mathcal{P}_\alpha,\, 1 \leq i \leq N \}.
\]
Define $\mathcal{X}^{(\alpha)} = \mathcal{Z}^{(\alpha)} \cap \{X_i\}_{i=1}^m$ and $\mathcal{Y}^{(\alpha)} = \mathcal{Z}^{(\alpha)} \cap \{Y_i\}_{i=1}^n$, so that $\mathcal{Z}^{(\alpha)}  = \mathcal{X}^{(\alpha)}  \cup \mathcal{Y}^{(\alpha)} $. Let $m_\alpha = |\mathcal{X}^{(\alpha)}|$, $n_\alpha = |\mathcal{Y}^{(\alpha)}|$, and $N_\alpha = |\mathcal{Z}^{(\alpha)}|$, where $|\cdot|$ denotes the cardinality of a set. Then
$$
\sum_{\alpha=1}^{n_P} m_\alpha = m, \quad \sum_{\alpha=1}^{n_P} n_\alpha = n, \quad \sum_{\alpha=1}^{n_P} N_\alpha = \sum_{\alpha=1}^{n_P} (m_\alpha + n_\alpha) = N.
$$
%During this stage, we partition our observation from both samples based on their missing patterns, which facilitates the implementation of Stage 2.

\paragraph{Stage 2:  Pattern-aware Procedure.} 
We then define a dissimilarity score for any two observations from the same or different missing patterns. For observations $Z_i \in \mathcal{Z}^{(\alpha)}$ and $Z_j \in \mathcal{Z}^{(\beta)}$, if their missing patterns share at least one modality (i.e., $\mathcal{P}_\alpha \mathcal{P}_\beta^\top > 0$), we define their dissimilarity as 
\begin{equation}\label{eq:naive distance}
\rho(Z_i, Z_j) = \mathtt{Norm} \Big(\sum_{l=1}^L S_i^{(l)} S_j^{(l)} \rho_l(Z_i^{(l)}, Z_j^{(l)}) \Big),
\end{equation}
where $\rho_l(\cdot, \cdot)$ is a chosen dissimilarity measure and $\mathtt{Norm}(\cdot)$ is a normalization function. Notice that the dissimilarity between a pair of observations only uses their shared modalities. For example, in Figure~\ref{fig:example_3sources}, given $Z_i \in \mathcal{Z}^{(3)}$ and $Z_j \in \mathcal{Z}^{(5)}$ with $\mathcal{P}_3 = (1,0,1)$ and $\mathcal{P}_5 = (1,0,0)$, the shared modality is $\mathcal{S}^{(1)}$, and their dissimilarity reduces to
$\rho(Z_i, Z_j)= \mathtt{Norm}\big(\rho_1(Z_i^{(1)},Z_j^{(1)})\big)$. If a pair of observations shares no common modalities, i.e., $\mathcal{P}_\alpha \mathcal{P}_\beta^\top = 0$, direct computation of dissimilarity is infeasible. We adopt a simple strategy by assigning zero values to such pairs, which performs well empirically in both simulations and real data analyses (see Sections~\ref{section: simluations} and~\ref{section: real data}). An alternative strategy that further exploits indirect comparability is discussed in Supplementary~M.

The dissimilarity in~\eqref{eq:naive distance} can take various forms depending on the relationships within and across modalities. For Euclidean modalities, one may use $\rho_l(Z_i^{(l)}, Z_j^{(l)}) = \|Z_i^{(l)} - Z_j^{(l)}\|_2^2$ with $\mathtt{Norm}(x) = \sqrt{x}$, yielding Euclidean distance over shared modalities. For non-Euclidean modalities, more flexible choices of $\rho_l(\cdot,\cdot)$ can be adopted, such as graph diffusion distances for networks \citep{Hammond2013}, cosine similarity for embeddings \citep{kenter2015short}, or tailored measures for categorical data \citep{boriah2008similarity}. Different choices of $\rho_l$ lead to different test-specific dissimilarities and test statistics.

Through Stage 2, pairwise dissimilarities are computed and organized into a block-structured dissimilarity matrix according to missingness patterns.

\paragraph{Stage 3: Test Statistic Assembly.} 

In Stage 3, we aggregate the block-structured dissimilarity matrix from Stage 2 into a global test statistic for testing $H_0$. Because different missingness patterns may involve varying sample sizes and shared modalities, direct application of standard two-sample methods to the full dissimilarity matrix may lead to invalid inference. To construct a valid test statistics, a possible strategy is to first compute pattern-wise/pattern-aware statistics based on the chosen testing method, and then aggregate information across missing patterns to construct a global test statistic. In Section~\ref{subsec: test statistic}, we follow this idea and propose new test statistics within the BPET framework using the RISE as the specific testing method. Other testing methods can also be adopted within the BPET framework.

\subsection{A Pattern-wise Permutation Procedure}

To conduct inference using the global test statistic, we approximate its null distribution via permutation. However, standard permutation preserves only the total sample sizes and ignores imbalance across missingness patterns, which can inflate type-I error under block-wise missingness. To address this issue, we propose a pattern-wise permutation scheme that permutes labels within each missingness pattern.

Let $Z=(Z^{(1)},\dots,Z^{(L)})$ denote the full data vector and $S\in\{0,1\}^L$ denote the missingness pattern. In two-sample testing, we have two groups/samples, i.e., Group X (samples in Group X are denoted by $X$) and Group Y (samples in Group Y are denoted by $Y$). Then for Group $G\in\{X,Y\}$, the full data follow $Z\sim F_G$, and the joint distribution of $(Z,S)$ is
\[
P_{Z, S\mid G}(z, s) = P_{S\mid Z, G}(s \mid z)\,\frac{dF_G(z)}{dz}.
\]
Under these notations, observed data consist of $(Z^{\mathrm{obs}},S)$, where $Z^{\mathrm{obs}}$ is the subvector of $Z$ indexed by $\{l:S^{(l)}=1\}$. Our inference is carried out under the null hypothesis $H_0: F_X=F_Y$, while allowing $P_{S\mid Z, G}(s \mid z)$ to differ between groups.

\begin{definition}[Permutation distributions]
\label{def:permdist}
Let $T=T(Z_1^{\mathrm{obs}},S_1,\dots,Z_N^{\mathrm{obs}},S_N;G_1,\dots,G_N)$ be a test statistic computed from the observed data $(Z_i^{\mathrm{obs}},S_i)$ with group labels $G_i\in\{X,Y\}$.

\emph{(i)} The \emph{standard permutation distribution} of $T$ is obtained by recomputing $T$ after uniformly permuting the labels $\{G_i\}$ over all $\binom{N}{m}$ assignments with $m$ labels $X$ and $n$ labels $Y$, keeping $\{(Z_i^{\mathrm{obs}},S_i)\}$ fixed. Under $H_0$, this is called the \emph{permutation null distribution}.

\emph{(ii)} The \emph{pattern-wise permutation distribution} is obtained by permuting the labels independently within each missingness pattern $\mathcal{P}_\alpha$, preserving $(m_\alpha, n_\alpha)$. Under $H_0$, this is called the \emph{pattern-wise permutation null distribution}.
\end{definition}

As an illustration of Definition~\ref{def:permdist}(ii), suppose there are two missingness patterns $\mathcal{P}_1$ and $\mathcal{P}_2$. If $\mathcal{P}_1$ contains $5$ observations with $3$ from Group $X$ and $2$ from Group $Y$, then the labels are permuted uniformly over all assignments preserving these counts. The same procedure is applied independently within $\mathcal{P}_2$, preserving $(m_\alpha,n_\alpha)$ for each pattern.

The proposed pattern-wise permutation is closely related to classical stratified permutation tests, where labels are permuted within predefined strata \citep{Fisher1935,LehmannRomano2005}. In our setting, the strata are defined by missingness patterns rather than fully observed covariates. Unlike classical stratification, missingness patterns may depend on unobserved components of the data, and the validity of such permutation schemes under block-wise missingness has not been previously established. In this work, we provide a sufficient condition (Theorem~\ref{ppstion-UB}) that guarantees the validity of pattern-wise permutation and, consequently, the proposed framework.

\subsection{Validity under Different Missing Assumptions}

We examine the validity of standard permutation under block-wise missingness and provide a sufficient condition under which pattern-wise permutation yields valid inference. We further show that this condition is strictly weaker than MCAR and can accommodate MNAR mechanisms.

Let $\widetilde T = T(Z_1^{\mathrm{obs}},S_1,\dots,Z_N^{\mathrm{obs}},S_N;\widetilde G_1,\dots,\widetilde G_N)$ with $\widetilde G_i\in\{X,Y\}$ denote the statistic recomputed under permutation. The following theorem establishes the failure of the standard permutation scheme when two groups/samples differ in their marginal missing pattern distributions.

\begin{theorem}[Failure of standard permutation under unequal missing pattern distributions]
\label{thm:stdperm-fail}
Suppose that for every pattern $s$ with positive probability under both groups, 
\[
F_X(\,\cdot\,\mid S=s) \;=\; F_Y(\,\cdot\,\mid S=s),
\]
but there exists at least one $s$ such that $P_{S\mid G=X}(s)\;\neq\; P_{S\mid G=Y}(s)$. Then under $H_0$, the distributions of $T$ and $\widetilde T$ differ when $\{\widetilde G_i\}$ are generated by the standard permutation scheme. Consequently, $p$-values based on the standard permutation distribution are invalid.
\end{theorem}

Theorem~\ref{thm:stdperm-fail} shows that standard permutation can fail whenever the two groups differ in their marginal distributions of missingness patterns, even if their conditional distributions within each pattern are identical. In such cases, rejection of $H_0$ may be driven by differences in missingness patterns rather than distributional differences between $F_X$ and $F_Y$, leading to inflated type-I error. Since unequal pattern distributions commonly arise in block-wise missing data due to heterogeneous data collection processes, standard permutation may be inappropriate in this setting.

\begin{theorem}[Validity of pattern-wise permutation]
\label{ppstion-UB}
Let $\widetilde T$ denote the statistic computed using pattern-wise permutation. Under $H_0: F_X = F_Y$, a sufficient condition for $\widetilde T \overset{d}{=} T$ is
\begin{equation}
\frac{P_{S\mid Z, G=X}(s \mid z)}{P_{S|G=X}(s)} = \frac{P_{S\mid Z, G=Y}(s \mid z)}{P_{S|G=Y}(s)} \quad \text{ for all } z,s \text{ with } P_{S|G}(s):=P_G(S=s)>0,
\label{cond1}
\end{equation}
where $G \in \{X,Y\}$. This condition is strictly weaker than MCAR, which requires
$P_{S\mid Z, G}(s \mid z) = P_{S\mid G}(s) \text{ a.s. } F_G \text{ for all } s$, $G \in \{X,Y\}$.
\end{theorem}

Condition~\eqref{cond1} is strictly weaker than MCAR and the assumptions required for standard permutation; whenever either holds, pattern-wise permutation is valid. Importantly, Condition~\eqref{cond1} allows the missingness mechanism to depend on the full data vector, including unobserved components, and thus accommodates MNAR mechanisms. This condition differs fundamentally from the MAR assumption, which restricts dependence to observed data only. Instead, Condition~\eqref{cond1} requires that the dependence of the missingness mechanism on the data is proportional across groups. To our knowledge, this is the first general criterion guaranteeing valid permutation inference under block-wise missingness. 

Under Condition~\eqref{cond1}, labels are exchangeable within each missingness pattern, and the pattern-wise permutation distribution coincides with the null distribution conditional on the observed patterns. Throughout the remainder of the paper, we assume that Condition~\eqref{cond1} holds.

\section{A Graph-Induced Rank-Based Test within BPET}
\label{subsec:NNG-induced Rank}

In this section, we incorporate the rank-weighted similarity graph method RISE into the proposed BPET framework to integrate modalities of heterogeneous types. RISE is a flexible graph-based two-sample testing approach that constructs test statistics using ranks induced from similarity graphs, such as the $k$-nearest neighbor graph ($k$-NNG) or the $k$-minimum spanning tree ($k$-MST), and naturally accommodates both Euclidean and non-Euclidean data geometries \citep{D.Zhou2023}. For simplicity, we focus on the $k$-NNG in this paper and extend the $k$-NNG-induced ranking scheme to block-wise missing data via pattern-aware graphs.

\subsection{Graph-induced Rank}

For a given $k$, the $k$-NNG is a directed graph in which each node represents an observation and each directed edge points from an observation to its $k$ nearest neighbors under a chosen dissimilarity measure. Based on the $k$-NNG, we define the induced rank as follows.

\begin{definition}[$k$-NNG-induced Rank]
\label{def:graph-induced-rank}
For any pair of observations $(Z_i, Z_j)$, the \emph{graph-induced rank} of $(Z_i, Z_j)$ is defined as
\[
\mathrm{rank}(Z_j \mid Z_i) = k + 1 - k',
\]
if $Z_j$ is the $k'$th nearest neighbor of $Z_i$ for $1 \leq k' \leq k$ and $\mathrm{rank}(Z_j \mid Z_i) = 0$ otherwise. When ties occur in the dissimilarity values, the tied neighbors are assigned the average of the ranks they would otherwise occupy.
\end{definition}

\begin{figure}[ht]
    \centering

    % Subfigure (a): Standard k-NNG with random node layout
    \begin{tikzpicture}[scale=1, 
        every node/.style={
        draw,
        circle,
        minimum size=5mm,      % Base size for all nodes
        inner sep=0.5pt,       % Reduce internal padding
        text width=3.5mm,      % Constrain text width
        align=center,          % Center text in constrained width
        font=\footnotesize,    % Smaller font if needed
        fill=white,            % White fill for nodes
        text=black             % Black text for numbers
    }]
        \node (1) at (0,0) {1};
        \node (2) at (1.5,1.2) {2};
        \node (3) at (3,0.2) {3};
        \node (4) at (2.3,2.4) {4};
        \node (5) at (0.8,2.5) {5};
        \node (6) at (-2,1.5) {6};

        % k=1 (blue)
        \draw[-Stealth, thick, blue] (1) to[bend left=10, looseness=0] (2);
        \draw[-Stealth, thick, blue] (2) to[bend left=10, looseness=0] (4);
        \draw[-Stealth, thick, blue] (3) to[bend left=10, looseness=0] (2);
        \draw[-Stealth, thick, blue] (4) to[bend left=10, looseness=0] (2);
        \draw[-Stealth, thick, blue] (5) to[bend left=10, looseness=0] (2);
        \draw[-Stealth, thick, blue] (6) to[bend left=10, looseness=0] (1);

        % k=2 (orange)
        \draw[-Stealth, thick, orange] (1) to[bend left=10, looseness=0] (6);
        \draw[-Stealth, thick, orange] (2) to[bend left=10, looseness=0] (5);
        \draw[-Stealth, thick, orange] (3) to[bend left=10, looseness=0] (4);
        \draw[-Stealth, thick, orange] (4) to[bend left=10, looseness=0] (5);
        \draw[-Stealth, thick, orange] (5) to[bend left=10, looseness=0] (4);
        \draw[-Stealth, thick, orange] (6) to[bend left=10, looseness=0] (5);

        % k=3 (green)
        \draw[-Stealth, thick, green!70!black] (1) to[bend left=10, looseness=0] (5);
        \draw[-Stealth, thick, green!70!black] (2) to[bend left=10, looseness=0] (3);
        \draw[-Stealth, thick, green!70!black] (3) to[bend left=10, looseness=0] (1);
        \draw[-Stealth, thick, green!70!black] (4) to[bend left=10, looseness=0] (3);
        \draw[-Stealth, thick, green!70!black] (5) to[bend left=10, looseness=0] (1);
        \draw[-Stealth, thick, green!70!black] (6) to[bend left=10, looseness=0] (2);
    \end{tikzpicture}
    \hspace{2cm}
    % Subfigure (b): Bipartite k-NNG with random node layout
    \begin{tikzpicture}[scale=1, 
        every node/.style={
        draw,
        circle,
        minimum size=5mm,      % Base size for all nodes
        inner sep=0.5pt,       % Reduce internal padding
        text width=3.5mm,      % Constrain text width
        align=center,          % Center text in constrained width
        font=\footnotesize     % Smaller font if needed
    }]
        % White nodes (white fill, black text)
        \node[fill=white, text=black] (a1) at (0,2.5) {1};
        \node[fill=white, text=black] (a2) at (0.3,1.2) {2};
        \node[fill=white, text=black] (a3) at (-0.3,0) {3};
        \node[fill=white, text=black] (a4) at (1.5,0.2) {4};

        % Black nodes (black fill, white text)
        \node[fill=olive, text=white] (b1) at (1.5,2.8) {5};
        \node[fill=olive, text=white] (b2) at (2.5,1.7) {6};
        \node[fill=olive, text=white] (b3) at (3,-0.3) {7};
        \node[fill=olive, text=white] (b4) at (0.2,-0.8) {8};

        % k = 1 (blue)
        \draw[-Stealth, thick, blue] (a1) to[bend left=10, looseness=0] (b1);
        \draw[-Stealth, thick, blue] (a2) to[bend left=10, looseness=0] (b1);
        \draw[-Stealth, thick, blue] (a3) to[bend left=10, looseness=0] (b4);
        \draw[-Stealth, thick, blue] (a4) to[bend left=10, looseness=0] (b3);

        % k = 2 (orange)
        \draw[-Stealth, thick, orange] (a1) to[bend left=10, looseness=0] (b2);
        \draw[-Stealth, thick, orange] (a2) to[bend left=10, looseness=0] (b2);
        \draw[-Stealth, thick, orange] (a3) to[bend left=10, looseness=0] (b3);
        \draw[-Stealth, thick, orange] (a4) to[bend left=10, looseness=0] (b4);

        % k = 3 (green)
        \draw[-Stealth, thick, green!70!black] (a1) to[bend right=10, looseness=0] (b4);
        \draw[-Stealth, thick, green!70!black] (a2) to[bend left=10, looseness=0] (b4);
        \draw[-Stealth, thick, green!70!black] (a3) to[bend left=10, looseness=0] (b2);
        \draw[-Stealth, thick, green!70!black] (a4) to[bend left=10, looseness=0] (b2);
    \end{tikzpicture}

    % Legend
    \vspace{0.5cm}
    \begin{tikzpicture}
        \draw[-Stealth, thick, blue] (0,0) -- (1,0);
        \node[anchor=west] at (1.2,0) {$k=1$};
        \draw[-Stealth, thick, orange] (3,0) -- (4,0);
        \node[anchor=west] at (4.2,0) {$k=2$};
        \draw[-Stealth, thick, green!70!black] (6,0) -- (7,0);
        \node[anchor=west] at (7.2,0) {$k=3$};
    \end{tikzpicture}

    \caption{Example of $k$th-NNGs for $(\mathcal{Z}^{(\alpha)},\mathcal{Z}^{(\beta)})$ on two cases: Left panel shows the case when $\alpha = \beta$;  Right panel shows the case when $\alpha \neq \beta$, inducing a bipartite graph. Colors of arrows indicate neighbor level, while colors of nodes indicate observations from different patterns.}
    \label{fig:kNNG-random}
\end{figure}
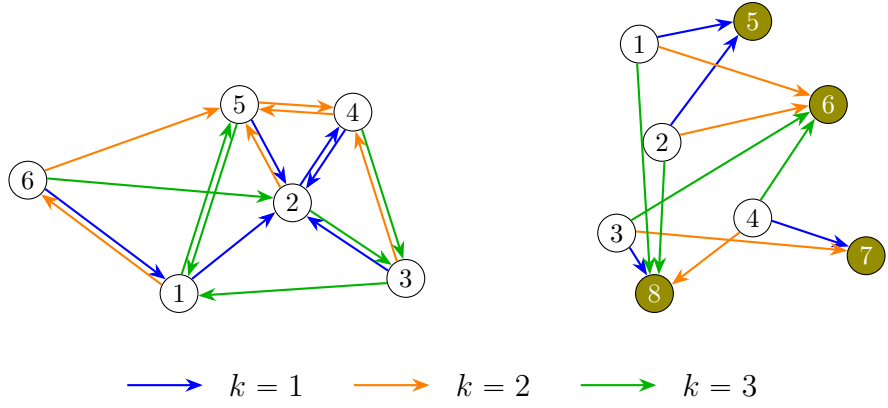

Now, we consider how to extend the $k$-NNG-induced rank to block-wise missing data. The key idea is to define a systematic ranking scheme that depends on the missingness patterns of each pair of observations. We first define pattern-aware $k$-NNGs. Given two observations $(Z_i, Z_j)$ belonging to missingness patterns $\mathcal{P}_\alpha$ and $\mathcal{P}_\beta$, respectively, the pattern-aware $k$-NNG is constructed using only observations from $\mathcal{Z}^{(\alpha)} \cup \mathcal{Z}^{(\beta)}$. In this graph, each node represents an observation in $\mathcal{Z}^{(\alpha)}$ or $\mathcal{Z}^{(\beta)}$, and each observation in $\mathcal{Z}^{(\alpha)}$ points to its $k$ nearest neighbors in $\mathcal{Z}^{(\beta)}$ via directed edges. An illustration is shown in Figure~\ref{fig:kNNG-random}. The structure of the pattern-aware $k$-NNG depends on whether $\mathcal{P}_\alpha = \mathcal{P}_\beta$:
\begin{itemize}
    \item If $\alpha = \beta$, then the pattern-aware $k$-NNG is a {standard $k$-NNG} with only observation from $\mathcal{P}_\alpha$ or $\mathcal{P}_\beta$, where bi-directional edges are possible (see the left panel in Figure~\ref{fig:kNNG-random});
    \item If $\alpha \neq \beta$, then the pattern-aware $k$-NNG is a {bipartite graph}, each observation in $\mathcal{Z}^{(\alpha)}$ points to its nearest neighbors in $\mathcal{Z}^{(\beta)}$, with no within-pattern edges (see the right panel in Figure~\ref{fig:kNNG-random}).
\end{itemize}
Then, for any pair of $(Z_i, Z_j)$ from missing patterns $(\mathcal{Z}^{(\alpha)}, \mathcal{Z}^{(\beta)})$, the \emph{rank of an observation} $Z_j$ w.r.t. $Z_i$ is the $k$-NNG-induced rank based on the pattern-aware $k$-NNG.

With the ranking scheme above, the rank of an observation $Z_j \in \mathcal{Z}^{(\beta)}$ with respect to $Z_i \in \mathcal{Z}^{(\alpha)}$ is denoted by $R_{ij}$. The corresponding \emph{total rank matrix} is $\mathbf{R} = [R_{ij}]_{i,j=1}^N$. Denote
\(
\mathbf{R}(\mathcal{W}^{(\alpha)}, \mathcal{W}^{(\beta)}) 
= [R_{ij} : Z_i \in \mathcal{W}^{(\alpha)},\; Z_j \in \mathcal{W}^{(\beta)}], 
\, \mathcal{W} \in \{\mathcal{X}, \mathcal{Y}, \mathcal{Z}\}.
\)
We can rearrange $\mathbf{R}$ into a block-wise matrix with each block corresponding to a pair of missing patterns within two samples/groups; for $(\mathcal{Z}^{(\alpha)}, \mathcal{Z}^{(\beta)})$, 
the corresponding \emph{pattern-pair rank matrix} (a sub-matrix of $\mathbf{R}$) is
\[
\mathbf{R}^{(\alpha\beta)} 
= \mathbf{R}(\mathcal{Z}^{(\alpha)}, \mathcal{Z}^{(\beta)}) = [R_{ij}^{(\alpha\beta)}]_{i,j=1}^{N_\alpha,N_\beta}:=
\begin{bmatrix}
    \mathbf{R}(\mathcal{X}^{(\alpha)}, \mathcal{X}^{(\beta)}) & 
    \mathbf{R}(\mathcal{X}^{(\alpha)}, \mathcal{Y}^{(\beta)}) \\
    \mathbf{R}(\mathcal{Y}^{(\alpha)}, \mathcal{X}^{(\beta)}) & 
    \mathbf{R}(\mathcal{Y}^{(\alpha)}, \mathcal{Y}^{(\beta)})
\end{bmatrix}.
\]
When $\mathcal{P}_\alpha \mathcal{P}_\beta^\top = 0$, i.e., no shared sources (or modalities), $\mathbf{R}^{(\alpha\beta)}$ would be a zero matrix.

\subsection{Test Statistics}
\label{subsec: test statistic}

We describe how to aggregate the pattern-pair rank matrices into global test statistics. For simplicity, we work with a symmetrized rank matrix $\mathbf{R} \leftarrow \tfrac{1}{2}(\mathbf{R} + \mathbf{R}^\top)$, which guarantees $R_{ij}^{(\alpha\beta)} = R_{ji}^{(\beta\alpha)}$. Throughout, $\mathbf{R}$ denotes this symmetrized matrix. For each pair of missing patterns $(\mathcal{Z}^{(\alpha)}, \mathcal{Z}^{(\beta)})$, define
\[
U_x^{(\alpha\beta)}= \sum_{i=1}^{m_\alpha} \sum_{j=1}^{m_\beta} R_{ij}^{(\alpha\beta)}, 
\quad  
U_y^{(\alpha\beta)} = \sum_{i=1}^{n_\alpha} \sum_{j=1}^{n_\beta} R_{ij}^{(\alpha\beta)},
\]
which represent the within-sample rank sums for Groups $X$ and $Y$.

An illustrative example in Supplementary~L shows that different alternatives may induce different deviation patterns: under scale or mixed alternatives, $U_x^{(\alpha\beta)}$ and $U_y^{(\alpha\beta)}$ can deviate in opposite directions for specific pattern pairs, whereas under location alternatives, they tend to shift in the same direction, with deviations accumulating across patterns.

Motivated by this observation, we propose two complementary aggregation strategies: a vectorized statistic sensitive to localized pattern-specific deviations, and a low-dimensional aggregated statistic capturing overall distributional shifts. Let
\[
\mathcal{I}=\{(\alpha,\beta):1\leq \alpha\leq\beta\leq n_P,\; \mathcal{P}_\alpha\mathcal{P}_\beta^\top>0\}.
\]
Define
\(
V = \big(U_x^{(\alpha\beta)}-\mu_x^{(\alpha\beta)},\; U_y^{(\alpha\beta)}-\mu_y^{(\alpha\beta)}\big)^\top_{(\alpha,\beta)\in\mathcal{I}}
\in \mathbb{R}^{2|\mathcal{I}|},
\)
where $\mu_x^{(\alpha\beta)}=\mathbb{E}(U_x^{(\alpha\beta)})$ and $\mu_y^{(\alpha\beta)}=\mathbb{E}(U_y^{(\alpha\beta)})$ are taken under the pattern-wise permutation null distribution. The vectorized Block-wise RISE statistic (BRISE-v) is
\begin{equation}
T_v=V^\top\boldsymbol{\Sigma}_v^{-1}V,
\end{equation}
with $\boldsymbol{\Sigma}_v=\mathrm{Cov}(V)$ under the same null. By construction, $T_v$ is sensitive to deviations arising from specific missing-pattern pairs.

Alternatively, define the aggregated rank sums
\[
{U}_x = \sum_{(\alpha,\beta) \in \mathcal{I}} U_x^{(\alpha\beta)} = \sum_{i=1}^m \sum_{j=1}^m R_{ij},
\qquad
{U}_y = \sum_{(\alpha,\beta) \in \mathcal{I}} U_y^{(\alpha\beta)} = \sum_{i=m+1}^N \sum_{j=m+1}^N R_{ij},
\]
which correspond to summing all within-group ranks. The congregated Block-wise RISE statistic (BRISE-c) is
\begin{equation}
\label{eq. BRISE-c}
T_c={\widebar U}^\top\boldsymbol{\Sigma}_c^{-1}{\widebar U},
\end{equation}
where $\widebar U = ({U}_x-\mu_x,{U}_y-\mu_y)^\top$, $\mu_x=\mathbb{E}({U}_x)$, $\mu_y=\mathbb{E}({U}_y)$, and $\boldsymbol{\Sigma}_c=\mathrm{Cov}(({U}_x,{U}_y)^\top)$. Compared to BRISE-v, BRISE-c aggregates information across all patterns and is therefore more sensitive to global distributional shifts.

The two statistics are complementary: BRISE-v targets heterogeneous local deviations across patterns, while BRISE-c emphasizes overall deviations. To avoid instability from extremely rare patterns when computing BRISE-v, we discard patterns with $\min\{m_\alpha,n_\alpha\}<n_{\mathrm{thres}}$ or $N_\alpha < p_{\mathrm{thres}} \max_\beta N_\beta$. Throughout, we use $n_{\mathrm{thres}}=2$ and $p_{\mathrm{thres}}=0.1$.

\begin{remark}
Other aggregation schemes are also possible. For example, weighted sums
\[
\widetilde{U}_x = \sum_{(\alpha,\beta) \in \mathcal{I}} a_x^{(\alpha\beta)} U_x^{(\alpha\beta)}, \qquad
\widetilde{U}_y = \sum_{(\alpha,\beta) \in \mathcal{I}} a_y^{(\alpha\beta)} U_y^{(\alpha\beta)}
\]
with prespecified weights could be substituted into \eqref{eq. BRISE-c}, or one could consider maximum-type statistics such as $\max_{(\alpha,\beta) \in \mathcal{I}, g \in \{X,Y\} }\{ Z_g^{\alpha \beta} \}$ where $Z_g^{(\alpha \beta)} = \frac{U_g^{(\alpha \beta)}-\mu_g^{(\alpha \beta)}}{\sigma_g^{(\alpha \beta)}}$ with $\sigma_g^{(\alpha \beta)}=\mathrm{Var}(U_g^{(\alpha \beta)})$. Exploration of such alternatives is left for future work.
\end{remark}

\begin{remark}
\label{rm: stability}
The covariance matrices $\boldsymbol{\Sigma}_v$ and $\boldsymbol{\Sigma}_c$ should be invertible to ensure $T_v$ and $T_c$ are well-defined. Simulations in Supplementary~N show that both are typically stable, with $\boldsymbol{\Sigma}_c$ more robust in small samples.  In rare cases of ill-conditioning, which occurs when some quantities $U_g^{(\alpha\beta)}$, $g \in \{x,y\}$, are linearly dependent, a subset of these terms can be removed to ensure invertibility of the covariance matrix.
\end{remark}

\section{Theoretical Properties}
\label{section: asymptotic properties}

\subsection{Finite-sample Properties}\label{subsec: fini prop}

In this section, we provide explicit formulas for $\boldsymbol{\Sigma}_{v}$ and $\boldsymbol{\Sigma}_{c}$ under the null hypothesis, which are crucial for constructing the proposed test statistics.

We first introduce notation related to the rank matrix. Let $\delta_{\alpha\beta} = \mathbbm{1}(\alpha = \beta)$ be the Kronecker delta with the indicator function $\mathbbm{1}(\cdot)$. We further define:
\begin{align*}
\widebar{R}_{i\cdot}^{(\alpha\beta)} &= 
\frac{1}{N_\beta-\delta_{\alpha\beta}}
\sum_{j=1}^{N_\beta}R_{ij}^{(\alpha\beta)},
&r_{0}^{(\alpha\beta)} &= \frac{1}{N_\alpha}\sum_{i=1}^{N_\alpha}
\widebar{R}_{i\cdot}^{(\alpha\beta)},\\
(r_1^{(\alpha\beta\gamma)})^2 &= 
\frac{1}{N_\alpha}\sum_{i=1}^{N_\alpha}
\widebar{R}_{i\cdot}^{(\alpha\beta)} 
\widebar{R}_{i\cdot}^{(\alpha\gamma)} ,
&(r_{2}^{(\alpha\beta)})^2 &= 
\frac{1}{N_\alpha(N_\beta-\delta_{\alpha\beta})}
\sum_{i=1}^{N_\alpha}\sum_{j=1}^{N_\beta}
({R}_{ij}^{(\alpha\beta)})^2,\\
V_1^{(\alpha\beta\gamma)} &=
(r_1^{(\alpha\beta\gamma)})^2
- r_{0}^{(\alpha\beta)}r_{0}^{(\alpha\gamma)},
&V_2^{(\alpha\beta)} &=
(r_{2}^{(\alpha\beta)})^2
-(r_{0}^{(\alpha\beta)})^2.
\end{align*}
Here $\widebar{R}_{i\cdot}^{(\alpha\beta)}$ denotes the average of the ranks in the $i$th row of $\mathbf R^{(\alpha\beta)}$; $r_{0}^{(\alpha\beta)}$ is the mean of $\widebar{R}_{i\cdot}^{(\alpha\beta)}$'s, which is also the average rank among $\mathbf R^{(\alpha\beta)}$; 
$(r_2^{(\alpha\beta)})^2$ and $(r_1^{(\alpha\beta\gamma)})^2$ are the second-order moments; $V_2^{(\alpha\beta)}$ and $V_1^{(\alpha\beta\gamma)}$ are two types of variance term corresponding to $r_2^{(\alpha\beta)}$ and $r_1^{(\alpha\beta\gamma)}$. We establish several useful properties of these quantities for any symmetric matrix $\mathbf{R}$ with zero diagonal; these properties simplify the calculation of $\boldsymbol{\Sigma}_{v}$ and $\boldsymbol{\Sigma}_{c}$. Details are given in Supplementary~C.

Let 
$
\mathcal{T}_x^{(\alpha,\beta)}=n_\alpha(m_{\beta}-1-\delta_{\alpha\beta}) (N_\beta-\delta_{\alpha\beta})
V_1^{(\alpha\beta\beta)},
$
$
\mathcal{T}_y^{(\alpha,\beta)}=m_\alpha(n_{\beta}-1-\delta_{\alpha\beta}) (N_\alpha-\delta_{\alpha\beta})
V_1^{(\beta\alpha\alpha)},
$ and
$
\mathcal{T}^{(\alpha,\beta)}=(N_\beta-\delta_{\alpha\beta})
V_1^{(\alpha\beta\beta)}
$. Theorem~\ref{Thm1} provides explicit expressions for the quantities used to construct $\boldsymbol{\Sigma}_{v}$ and $\boldsymbol{\Sigma}_{c}$.

\begin{theorem}\label{Thm1}
Under the pattern-wise permutation null distribution, for $1\leq \alpha,\beta,\gamma\leq n_P$ and $\beta\neq\gamma$, with $\min_{1\leq \alpha\leq n_P}\{m_\alpha,n_\alpha\}\geq 2$, we have
$$
    \mu_x^{(\alpha\beta)}=\mathbb{E}(U_x^{(\alpha\beta)})=
    m_\alpha(m_\beta-\delta_{\alpha\beta})
    r_{0}^{(\alpha\beta)}, \quad
    \mu_y^{(\alpha\beta)}=\mathbb{E}(U_y^{(\alpha\beta)})=
    n_\alpha(n_\beta-\delta_{\alpha\beta})
    r_{0}^{(\alpha\beta)},
$$
$$
\mathrm{Var}(U_x^{(\alpha\beta)}) = 
\frac{
(1+\delta_{\alpha\beta})m_\alpha (m_\beta-\delta_{\alpha\beta})
}{(N_\alpha-1-\delta_{\alpha\beta})(N_\beta-1-2\delta_{\alpha\beta})}
\Big(
{n_\alpha (n_\beta-\delta_{\alpha\beta})}
V_2^{(\alpha\beta)}
+
\mathcal{T}_x^{(\alpha,\beta)}
+
\mathcal{T}_x^{(\beta,\alpha)}
\Big),
$$
$$
\mathrm{Var}(U_y^{(\alpha\beta)}) = 
\frac{
(1+\delta_{\alpha\beta})n_\alpha (n_\beta-\delta_{\alpha\beta})
}{(N_\alpha-1-\delta_{\alpha\beta})(N_\beta-1-2\delta_{\alpha\beta})}
\Big(
{m_\alpha (m_\beta-\delta_{\alpha\beta})}
V_2^{(\alpha\beta)}
+
\mathcal{T}_y^{(\alpha,\beta)}
+
\mathcal{T}_y^{(\beta,\alpha)}
\Big),
$$
$$
\mathrm{Cov}(U_x^{(\alpha\beta)},U_y^{(\alpha\beta)}) = 
\frac{
(1+\delta_{\alpha\beta})m_\alpha 
n_\alpha 
(m_\beta-\delta_{\alpha\beta})
(n_\beta-\delta_{\alpha\beta})
}{(N_\alpha-1-\delta_{\alpha\beta})(N_\beta-1-2\delta_{\alpha\beta})}
\Big(
V_2^{(\alpha\beta)}
- 
\mathcal{T}^{(\alpha,\beta)}
- 
\mathcal{T}^{(\beta,\alpha)}
\Big),
$$
$$
    \mathrm{Cov}(U_x^{(\alpha\beta)},U_x^{(\alpha\gamma)}) =
    \frac{(1+\delta_{\alpha\beta}+\delta_{\alpha\gamma})m_\alpha n_\alpha (m_\beta-\delta_{\alpha\beta}) (m_\gamma-\delta_{\alpha\gamma}) }{N_\alpha-1-\delta_{\alpha\beta}-\delta_{\alpha\gamma}}
    V_1^{(\alpha\beta\gamma)},
$$
$$
    \mathrm{Cov}(U_y^{(\alpha\beta)},U_y^{(\alpha\gamma)}) =
    \frac{(1+\delta_{\alpha\beta}+\delta_{\alpha\gamma})m_\alpha n_\alpha (n_\beta-\delta_{\alpha\beta}) (n_\gamma-\delta_{\alpha\gamma}) }{N_\alpha-1-\delta_{\alpha\beta}-\delta_{\alpha\gamma}}
    V_1^{(\alpha\beta\gamma)},
$$
$$
    \mathrm{Cov}(U_x^{(\alpha\beta)},U_y^{(\alpha\gamma)}) =
    -\frac{(1+\delta_{\alpha\beta}+\delta_{\alpha\gamma})m_\alpha n_\alpha (m_\beta-\delta_{\alpha\beta}) (n_\gamma-\delta_{\alpha\gamma}) }{N_\alpha-1-\delta_{\alpha\beta}-\delta_{\alpha\gamma}}
    V_1^{(\alpha\beta\gamma)}.
$$
In cases other than above, the covariances are equal to 0.

\end{theorem}

Using the result in Theorem~\ref{Thm1}, we can directly calculate the BRISE-v statistic. While for calculating BRISE-c, we further utilize the following relations:
$$
    \mu_x=\mathbb{E}({U_x}) = 
    \sum_{(\alpha,\beta) \in \mathcal{I}} 
    \mathbb{E}(U_x^{(\alpha\beta)})
    ,\quad
    \mu_y=\mathbb{E}({U_y}) = 
     \sum_{(\alpha,\beta) \in \mathcal{I}} 
    \mathbb{E}(U_y^{(\alpha\beta)}), 
$$
$$
\mathrm{Var}({U_x}) = 
\sum_{(\alpha,\beta) \in \mathcal{I}} 
\sum_{(\gamma,\omega) \in \mathcal{I}} 
\mathrm{Cov}(U_x^{(\alpha\beta)},U_x^{(\gamma\omega)}), \,
\mathrm{Var}({U_y}) = 
\sum_{(\alpha,\beta) \in \mathcal{I}} 
\sum_{(\gamma,\omega) \in \mathcal{I}} 
\mathrm{Cov}(U_y^{(\alpha\beta)},U_y^{(\gamma\omega)}),
$$
$$
\mathrm{Cov}({U_x},{U_y}) =
\sum_{(\alpha,\beta) \in \mathcal{I}} 
\sum_{(\gamma,\omega) \in \mathcal{I}} 
\mathrm{Cov}(U_x^{(\alpha\beta)},U_y^{(\gamma\omega)}).
$$
In particular, when $\alpha=\beta$, these moments reduce to those in Theorem~1 of \cite{D.Zhou2023}, corresponding to the fully observed setting. Thus, RISE can be viewed as a special case of the proposed framework with fully observed data.

\subsection{Asymptotic Properties}

Exact $p$-value computation via full pattern-wise permutation is feasible for small samples but becomes computationally intractable as sample size grows. To enable scalable inference, we derive the asymptotic distributions of the test statistics $T_v$ and $T_c$ under the null. We use the following asymptotic notations: $a_n \prec b_n$ indicates $a_n$ is dominated by $b_n$; $a_n \asymp b_n$ denotes boundedness above and below; and $a_n \precsim b_n$ denotes boundedness above. We adopt a standard asymptotic regime where $m_\alpha, n_\alpha \to \infty$, with $\lim_{N \to \infty} N_\alpha/N \in (0,1)$ and $m_\alpha / N_\alpha \to p_\alpha \in (0,1)$ for all $1 \leq \alpha \leq n_P$.

To introduce our main result, we further define the following notations. For each pattern pair $(\mathcal{Z}^{(\alpha)}, \mathcal{Z}^{(\beta)})$, define the centered row sum $\widetilde{R}_{i\cdot}^{(\alpha\beta)} 
= \widebar{R}_{i\cdot}^{(\alpha\beta)} - r_0^{(\alpha\beta)}$.
To capture the cross-pattern second-order dependence, we define
\[
\mathbf{E_\alpha} = [\epsilon_{\alpha \beta \gamma}]_{\beta\gamma}, \text{ where }  
\epsilon_{\alpha\beta\gamma} 
= (r_1^{(\alpha\beta\gamma)})^2 
- \frac{(r_1^{(\alpha\alpha\beta)})^2 (r_1^{(\alpha\alpha\gamma)})^2}{(r_1^{(\alpha\alpha\alpha)})^2} \text{ for }(\alpha,\beta), (\alpha,\gamma) \in \mathcal{I}, \beta \neq \alpha, \gamma \neq \alpha .
\]

\begin{theorem}[Limiting distribution under the null hypothesis]
\label{Thm2}
In the standard limit regime, for all $(\alpha,\beta), (\beta,\gamma) \in \mathcal{I}$, under Conditions
$(1)\ r_1^{(\alpha\beta\beta)} \prec r_2^{(\alpha\beta)}$, $
(2)\ \sum_{i=1}^{N_\alpha}\!\left(\sum_{j=1}^{N_\beta}(R_{ij}^{(\alpha\beta)})^2\right)^2
\lesssim N_\alpha^2 N_\beta (r_2^{(\alpha\beta)})^4$, $
(3)\ \sum_{i=1}^{N_\alpha}\lvert\widetilde R_{i\cdot}^{(\alpha\beta)}\rvert^3
\prec (N_\alpha V_1^{(\alpha\beta\beta)})^{3/2}$,
$(4)\ \sum_{i=1}^{N_\alpha}\lvert\widetilde R_{i\cdot}^{(\alpha\beta)}\rvert^3
\prec N_\alpha r_2^{(\alpha\beta)} V_1^{(\alpha\beta\beta)}$, \\
$(5)\ \Bigl|\sum_{i=1}^{N_\alpha}\sum_{j\neq l}^{N_\beta}
R_{ij}^{(\alpha\beta)}R_{il}^{(\alpha\beta)}
\widetilde R_{j\cdot}^{(\beta\gamma)}\widetilde R_{l\cdot}^{(\beta\gamma)}\Bigr|
\prec N_\alpha N_\beta^2 (r_2^{(\alpha\beta)})^2 V_1^{(\beta\gamma\gamma)}$,
$(6)\ \sum_{i\neq l}^{N_\alpha}\sum_{j\neq s}^{N_\beta}
R_{ij}^{(\alpha\beta)}R_{lj}^{(\alpha\beta)}R_{is}^{(\alpha\beta)}R_{ls}^{(\alpha\beta)}
\prec N_\alpha^2 N_\beta^2 (r_2^{(\alpha\beta)})^4$
and (C1): $\mathbf{E_\alpha}$ is positive definite, 
we have 
$$T_v\overset{\mathcal{D}}{\to}\chi_a^2 \text{ and }  T_c\overset{\mathcal{D}}{\to}\chi_2^2$$ under the pattern-wise permutation null distribution, 
where $a = 2|\mathcal{I}|$ and $\overset{\mathcal{D}}{\to}$ is convergence in distribution. 
\end{theorem}

The validity of Conditions~(1)--(6) depends on the indices $\alpha$, $\beta$, and $\gamma$. When $\alpha=\beta=\gamma$, these conditions reduce to those in Theorem~4 of \cite{D.Zhou2023}, which were shown to be mild and plausible in practice. When $\alpha,\beta,\gamma$ are not all the same, Conditions (1), (2), (4), and (6) hold assuming that there are no nodes with a degree larger than $Ck$ for some constant $C$ in the pair-wise $k$-NNG defined in Section~\ref{subsec:NNG-induced Rank}. To see this, let $G^{(\alpha\beta)}$ be the $k$-NNG given a pair of missing patterns $(\mathcal{Z}^{(\alpha)},\mathcal{Z}^{(\beta)})$, then we have $k = \max_{\alpha\beta,ij} R_{ij}^{(\alpha\beta)}$, and adopt the assumption that $k\prec N$ for the graph-induced rank. With a slight abuse of notation, we use $|\cdot|$ to denote the total amount of edges in a graph. Usually we have $r_0^{(\alpha\beta)} \asymp k|G^{(\alpha\beta)}|/N_\alpha N_\beta$ and $(r_2^{(\alpha\beta)})^2 \asymp k^2|G^{(\alpha\beta)}|/N_\alpha N_\beta$ where $|G^{(\alpha\beta)}| \asymp (N_\alpha+N_\beta) k/2$. Thus, we have that Conditions (1), (2), (4), and (6) always hold as
$$
(r_1^{(\alpha\beta\beta)})^2
= \frac{1}{N_\alpha N_\beta^2}
\sum_{i=1}^{N_\alpha}
(\sum_{j=1}^{N_\beta}
R_{ij}^{(\alpha\beta)})^2
\asymp \frac{k^4}{N_\beta^2}
\prec({r_2^{(\alpha\beta)}})^2,
$$
$$
{\sum_{i=1}^{N_\alpha}(\sum_{j=1}^{N_\beta} (R_{ij} ^{(\alpha\beta)})^2)^2}
\asymp N_\alpha (k^3)^2
\asymp N_\alpha (N_\alpha+N_\beta)^2 ({r_2^{(\alpha\beta)}})^4
\asymp{N_\alpha^2 N_\beta (r_2^{(\alpha\beta)})^4},
$$
$$
\sum_{i=1}^{N_\alpha}
{|\widetilde R_{i\cdot}^{(\alpha\beta)}|^3}
\leq \max_i |\widetilde R_{i\cdot}^{(\alpha\beta)}|N_\alpha V_1^{(\alpha\beta\beta)}
\lesssim \frac{k^2}{N_\alpha} N_\alpha V_1^{(\alpha\beta\beta)}
\prec N_\alpha r_2^{(\alpha\beta)} V_1^{(\alpha\beta\beta)},
$$
$$
\begin{aligned}
&
\sum_{i=1}^{N_\alpha}
\sum_{j=1}^{N_\beta}
\sum_{l=1, l\neq i}^{N_\alpha}
\sum_{s=1, s\neq j}^{N_\beta}
R_{ij} ^{(\alpha\beta)}
R_{lj} ^{(\alpha\beta)}
R_{is} ^{(\alpha\beta)}
R_{ls} ^{(\alpha\beta)}
\lesssim
\sum_{i=1}^{N_\alpha}
\sum_{j=1}^{N_\beta}
R_{ij} ^{(\alpha\beta)}
\sum_{l=1}^{N_\alpha}
R_{lj} ^{(\alpha\beta)}
\sum_{s=1}^{N_\beta}
R_{is} ^{(\alpha\beta)}
R_{ls} ^{(\alpha\beta)}\\
&\lesssim
k
\sum_{i=1}^{N_\alpha}
\sum_{j=1}^{N_\beta}
R_{ij} ^{(\alpha\beta)}
\sum_{l=1}^{N_\alpha}
R_{lj} ^{(\alpha\beta)}
\sum_{s=1}^{N_\beta}
R_{is} ^{(\alpha\beta)}
\asymp
N_\alpha k^7
\asymp N_\alpha (N_\alpha+N_\beta)^2 k (r_2^{(\alpha\beta)})^4
\prec N_\alpha^2N_\beta^2(r_2^{(\alpha\beta)})^4.
\end{aligned}
$$
We further verify Conditions (3) and (5) via simulation by evaluating the ratio of the left-hand side to the right-hand side of each inequality. Together with Condition (C1), simulation details and results are summarized in Supplementary~O. The simulation results support that these conditions are easily satisfied.

We next establish the consistency of BRISE. For the consistency under the standard limiting regime (see Theorem~\ref{Thm3: Consistency}), Conditions (1) -~(6) are not required. We consider $G^{(\alpha\beta)}$ constructed using Euclidean distance. When $k = O(1)$ and $N_\alpha, N_\beta \to \infty$, each edge in $G^{(\alpha\beta)}$ exhibits asymptotically vanishing length, with connected nodes converging to arbitrarily close proximity in space \citep{Henze1999}. This property arises because, in the limit, any point possesses sufficiently many neighbors within an arbitrarily small neighborhood to constitute its nearest neighbors. Under the choice of Euclidean distance, we have the following theorem.

\begin{theorem}[Consistency]
\label{Thm3: Consistency}
For two continuous multivariate distributions $F_X$ and $F_Y$ differing on a set of positive Lebesgue measure, employing the graph-induced rank with the $k$-NNG based on Euclidean distance, where $k = O(1)$, the power of both \emph{BRISE-v} and \emph{BRISE-c} at level $\theta \in (0,1)$ approaches $1$ in the standard limiting regime.
\end{theorem}

The proof of Theorem~\ref{Thm3: Consistency} is detailed in Supplementary~F, following the methodologies of \cite{schilling1986multivariate} and \cite{Henze1999}. This result may extend to alternative dissimilarity measures, provided they exhibit the same limiting behavior as $G^{(\alpha\beta)}$.

While Theorem~\ref{Thm3: Consistency} establishes the consistency of BRISE in the standard asymptotic settings, many modern applications, like genomics, neuroimaging, and text data, operate in the high-dimensional, low-sample-size (HDLSS) regime. In this context, classical asymptotic theory may fail due to the curse of dimensionality and the degeneracy of distance metrics. The following result extends the consistency of BRISE to the HDLSS setting under mild moment and separation conditions.

\begin{theorem}[Consistency under HDLSS]
\label{Thm4: HDLSS}
Assume that $F_X$ and $F_Y$ satisfy the following assumptions:

\emph{(i)} There exist
$\sigma_{x,\alpha\beta}^2, \sigma_{y,\alpha\beta}^2 > 0$ and $\nu_{\alpha\beta}^2 > 0$
such that for $X^{(\alpha)} \in \mathcal{X}^{(\alpha)}$, $X^{(\beta)} \in \mathcal{X}^{(\beta)}$, $Y^{(\alpha)} \in \mathcal{Y}^{(\alpha)}$, and $Y^{(\beta)} \in \mathcal{Y}^{(\beta)}$ sampled independently from $F_X$ and $F_Y$, for $(\alpha,\beta)\in\mathcal{I}$, the following limits hold:
$$
\frac{1}{d_{\alpha\beta}} \rho(\mathbb{E}(\rho((X^{(\alpha)},\mathbb{E}X^{(\beta)})^2)\to \sigma_{x,\alpha\beta}^2, \qquad
\frac{1}{d_{\alpha\beta}} \mathbb{E}(\rho(Y^{(\alpha)}, \mathbb{E}Y^{(\beta)})^2) \to \sigma_{y,\alpha\beta}^2,$$
$$
\frac{1}{d_{\alpha\beta}} \mathbb{E}(\rho(\mathbb{E}X^{(\alpha)},\mathbb{E}Y^{(\beta)})^2) \to \nu_{\alpha\beta}^2,
$$
where $d_{\alpha\beta}$ denotes the dimension of the overlapping sources for pattern pair $(\mathcal{Z}_\alpha,\mathcal{Z}_\beta)$ and here $\rho$ is the distance measure define in \eqref{eq:naive distance} with $\rho_l$ being the squared Euclidean distance and $\mathtt{Norm}(x) = \sqrt{x}$.

\emph{(ii)} The fourth moments of the components of all observed sources are uniformly bounded.

\emph{(iii)} 
Let $Z_{iq}$ be the $q$th variable of a given observation $Z_i$. For $(Z_{iq}, Z_{jq})_{Z_i\sim F_g, Z_j\sim F_{g'}}$ where ${g, g'\in \{X,Y\}}$, we assume the sequence $\{(Z_{iq}, Z_{jq}),\;q\geq 1\}$ is $\rho'$-mixing, i.e.,
\[
\sup_{1 \leq q < q' < \infty, \, |q-q'|>r} 
\big| \mathrm{corr}\{(Z_{iq}-Z_{jq})^2, (Z_{iq'}-Z_{jq'})^2\} \big| \leq \rho'(r),
\]
with $\rho'(r) \to 0$ as $r \to \infty$.

For any fixed significance level $\theta \in (0,1)$, we have
$$
\lim_{d \to \infty} {P}\left(T_v > \chi^2_a(1 - \theta)\right) = 1 \quad\text{and}\quad
\lim_{d \to \infty} {P}\left(T_c > \chi^2_2(1 - \theta)\right) = 1 ,
$$
where the probability here is under the alternative hypothesis if the alternative hypothesis satisfies the following conditions:
\begin{enumerate}
    \item $k < \min_{1\leq \alpha\leq n_P}\{n_\alpha, m_\alpha\},$ $\lim_{d \to \infty} {d_{\alpha\beta}}/{d} \in (0,1)$ for all $1 \leq \alpha \leq \beta \leq n_P$;
    \item There exists at least one $(\alpha,\beta)$ such that either $\sigma_{x,\alpha\beta}^2 \ne \sigma_{y,\alpha\beta}^2$ or $\nu_{\alpha\beta}^2 > 0$;
    \item The largest eigenvalue of the covariance matrix $\boldsymbol{\Sigma}_v$, denoted $\lambda_+[\boldsymbol{\Sigma}_v]$, is bounded by $\lambda_+[\boldsymbol{\Sigma}_v] \leq cN^{2 - b}$ for constants $c,b > 0$, and $N\geq C_{b,\theta}$ where $C_{b,\theta} > 0$ is a constant only depending on $b,\theta$.
\end{enumerate}
\end{theorem}

Assumptions (i) -~(iii) of Theorem~\ref{Thm4: HDLSS} are extended from \cite{biswas2014nonparametric}. Here we focus on the overlapping Euclidean distance case in \eqref{eq:naive distance}. The proposed tests may remain consistent under other dissimilarity measures, provided analogous limiting conditions hold.

\section{Simulation Studies}
\label{section: simluations}

We conduct simulation studies to evaluate the validity of BPET framework, and empirical type-I error and power of BRISE. We consider both BRISE-c and BRISE-v with Euclidean distance in \eqref{eq:naive distance} and $k=10$ for constructing the $k$-NNG, following the recommendation of \citet{D.Zhou2023}. We also examine the sensitivity of our methods to the choice of $k$ (see additional results reported in  Supplementary~P). The findings indicate that the performance of BRISE is stable with respect to $k$, as long as $k$ is not chosen too small. For both variants, $p$-values are obtained using their asymptotic null distributions.

We compare our methods with five baselines: MMD-Miss, RISE with $10$-NNG, the standard MMD test \citep{MMD}, Ball Divergence (BD) \citep{BD}, and Measure Transportation (MT) \citep{MT}, with their default hyperparameters. Among these, only MMD-Miss can be directly applied to incomplete data; the others are implemented via complete-case analysis. The $p$-values of MMD-Miss, MMD, MT, and BD are obtained using $1000$ permutations, while RISE uses its asymptotic null distribution for $p$-value calculation.

To further assess whether imputation can serve as a viable alternative under block-wise missingness, we also consider an imputation-based strategy in low-dimensional settings: we first apply MICE \citep{VanBuurenGroothuisOudshoorn2011} to impute the missing data, and then apply RISE, MMD, BD, and MT, respectively, to the imputed datasets. This comparison is restricted to $d \in \{10,20,50\}$ due to the computational cost and instability of multivariate imputation in higher dimensions.

To examine the advantage of the proposed pattern-wise permutation procedure in type-I error control, we further include a variant of BRISE-c that differs only in the permutation scheme, using standard instead of pattern-wise permutation, denoted as BRISE-c(sp). According to Theorem~\ref{thm:stdperm-fail}, this variant may fail when the marginal pattern probabilities differ between groups. We emphasize that most competing methods are not designed for block-wise missingness and therefore require either complete-case analysis or model-based imputation. Our goal is not to advocate a particular imputation model, but to provide a test that operates directly on observed data without imputation.

To simulate block-wise missing data, for each observation, each source/modality $Z^{(l)} \in \mathbb{R}^{d_l}$ is generated and observed independently with probability $p_{G}$ for $G \in \{X, Y\}$; observations with all sources missing are discarded. In total, we have \(d = \sum_{l=1}^L d_l\) variables.  To reflect a variety of realistic scenarios, we simulate a wide range of null and alternative distributions in moderate- to high-dimensional settings, including  multivariate Gaussian, log-normal, and $t_5$ distributions, covering both light-tailed and heavy-tailed behaviors. The alternatives include location, scale, and mixed differences. Specifically, we consider the following source/modality distributions:

\begin{enumerate}[label=(\Roman*)]
    \item \textbf{Multivariate Gaussian.}  
    $(X^{(1)},\ldots, X^{(L)}) \sim \mathcal{N}_{d}(\mathbf 0_{d}, \Sigma_X)$, where $\Sigma_{X,ij} = 0.6^{|i-j|}$. $(Y^{(1)},\ldots, Y^{(L)}) \sim F_Y(\cdot)$, where $F_Y(\cdot)$ is:
    \begin{enumerate}[label=\alph*.]
        \item \textit{Location:}  
        $\mathcal{N}_{d}(\boldsymbol \mu, \Sigma_X)$, where $\boldsymbol \mu = 0.4 \log {d} \cdot \boldsymbol \mu' / \|\boldsymbol \mu'\|_2$ and $\boldsymbol \mu' \sim \mathcal{N}_d(0_{d}, I_{d})$;
        
        \item \textit{Scale:}  
        $\mathcal{N}_{d}(\mathbf 0_{d}, \sigma^2 \Sigma_X)$, with $\sigma = 1 + 0.12 \log {d} / \sqrt{{d}}$;
        
        \item \textit{Mixed:}  
        $\mathcal{N}_{d}(\boldsymbol \mu, \Sigma_Y)$, where $\boldsymbol \mu = 0.1 \log {d} \cdot \boldsymbol \mu'/\|\boldsymbol \mu'\|_2$, $\boldsymbol \mu' \sim \mathcal{N}_d(0_{d}, I_{d})$, and $\Sigma_{Y,ij} = 0.3^{|i-j|}$.
    \end{enumerate}

    \item \textbf{Multivariate Log-Normal.}  
     $(X^{(1)},\ldots, X^{(L)}) \sim \exp(\mathcal{N}_{d}(\mathbf 0_{d}, \Sigma_X))$, with $\Sigma_{X,ij} = 0.6^{|i-j|}$. $(Y^{(1)},\ldots, Y^{(L)}) \sim F_Y(\cdot)$, where $F_Y(\cdot)$ is:
    \begin{enumerate}[label=\alph*.]
        \item \textit{Location:}  
        $\exp(\mathcal{N}_{d}(\boldsymbol \mu, \Sigma_X))$, where $\mu_j = (-1)^j \cdot 2 \log {d} / \sqrt{{d}}$ for $j = 1, \dots, \lfloor 0.05{d} \rfloor$, and $\mu_j = 0$ otherwise;

        \item \textit{Scale:}  
        $\exp(\mathcal{N}_{d}(\mathbf 0_{d}, \sigma^2 \Sigma_X))$, with $\sigma = 1 + 0.15 \log {d} / \sqrt{{d}}$;

        \item \textit{Mixed:}  
        $\exp(\mathcal{N}_d(\mu \mathbf 1_{d}, \sigma \Sigma_X))$, where $\mu = 0.25 \log {d} / \sqrt{{d}}$ and $\sigma = 1 + 0.1 \cdot (50/{d})^{0.25}$.
    \end{enumerate}

    \item \textbf{Multivariate $t_5$.}  
    $(X^{(1)},\ldots, X^{(L)}) \sim t_5(\mathbf 0_{d}, \Sigma_X)$, where $\Sigma_{X,ij} = 0.6^{|i-j|}$. $(Y^{(1)},\ldots, Y^{(L)}) \sim F_Y(\cdot)$, where $F_Y(\cdot)$ is:
    \begin{enumerate}[label=\alph*.]
        \item \textit{Location:}  
        $t_5(\boldsymbol \mu, \Sigma_X)$, where $\mu_j = (-1)^j \cdot 2 \log {d} / \sqrt{{d}}$ for $j = 1, \dots, \lfloor 0.05{d} \rfloor$, and $\mu_j = 0$ otherwise;

        \item \textit{Scale:}  
        $t_5(\mathbf 0_{d}, \Sigma_Y)$, where $\Sigma_{Y,ij} = 0.75 \cdot 0.3^{|i-j|}$;

        \item \textit{Mixed:}  
        $t_5(\mu \mathbf 1_{d}, \Sigma_Y)$, with $\mu = 0.4 \log {d} / \sqrt{{d}}$ and $\Sigma_{Y,ij} = 0.8^{|i-j|}$.
    \end{enumerate}
\end{enumerate}

We set the default setting as $m = n = 100$, $L = 2$, $p_X = p_Y = 0.5$, and $d \in \{200, 500, 1000\}$. We also include additional experiments in Supplementary~P, including imbalanced sample sizes ($m=200, n=100$), an increased number of sources ($L=3$), varied sampling rates ($p_X = p_Y \in \{0.2, 0.8\}$), and imbalanced sampling rates ($p_X=0.5, p_Y = 0.8$). We repeat simulations $1000$ times for empirical sizes and power.

At significance level $\theta = 0.05$, the empirical sizes under both balanced and imbalanced sampling rates are reported in Table~\ref{sim: Empirical sizes2}. When the sampling rates are balanced, all methods exhibit acceptable type-I error. When the sampling rates are imbalanced, BRISE-c, BRISE-v, MMD-Miss, and the complete-case methods still control type-I error well, whereas BRISE-c(sp) fails to do so. This highlights the importance of accounting for block-wise missingness via pattern-wise permutation. 

In the low-dimensional experiments, the imputation-based methods using MICE fail to control the type-I error, despite being computationally feasible in these settings. This suggests that imputation may not provide valid inference under block-wise missingness, even when dimensionality is low.

\begin{table}[htbp]
    \centering
    \caption{Empirical sizes of the tests at significance level $\theta = 0.05$, for $m = n = 100$, $L = 2 $, repeated 1000 times. Here we use \textit{italic} type to denote values way higher than $0.05$.}
    {\footnotesize
    \begin{tabular}{|c|ccc|ccc|ccc|}
        \hline
        $d$ & 200 & 500 & 1000 & 200 & 500 & 1000 & 200 & 500 & 1000 \\
        \hline
        $ p_X = p_Y = 0.5$ & \multicolumn{3}{c|}{Setting I} & \multicolumn{3}{c|}{Setting II} & \multicolumn{3}{c|}{Setting III} \\
        \hline
        BRISE-c        & 0.035 & 0.041 & 0.065 & 0.039 & 0.049 & 0.061 & 0.056 & 0.047 & 0.043 \\
        BRISE-v        & 0.031 & 0.045 & 0.049 & 0.055 & 0.050 & 0.047 & 0.056 & 0.051 & 0.043 \\
        MMD-Miss       & 0.000 & 0.000 & 0.000 & 0.000 & 0.000 & 0.000 & 0.000 & 0.000 & 0.000 \\
        RISE           & 0.041 & 0.040 & 0.060 & 0.048 & 0.042 & 0.055 & 0.054 & 0.057 & 0.039 \\
        MMD            & 0.000 & 0.000 & 0.000 & 0.002 & 0.000 & 0.000 & 0.005 & 0.004 & 0.002 \\
        BD             & 0.045 & 0.040 & 0.050 & 0.052 & 0.043 & 0.057 & 0.052 & 0.041 & 0.039 \\
        MT             & 0.044 & 0.043 & 0.057 & 0.042 & 0.040 & 0.061 & 0.047 & 0.035 & 0.039 \\
        BRISE-c(sp)       & 0.034 & 0.048 & 0.064 & 0.045 & 0.051 & 0.058 & 0.045 & 0.048 & 0.046 \\
        \hline
        $ p_X = 0.5, p_Y = 0.8$ & \multicolumn{3}{c|}{Setting I} & \multicolumn{3}{c|}{Setting II} & \multicolumn{3}{c|}{Setting III} \\
        \hline
        BRISE-c        & 0.037 & 0.047 & 0.064 & 0.040 & 0.044 & 0.046 & 0.043 & 0.040 & 0.049 \\
        BRISE-v        & 0.040 & 0.042 & 0.054 & 0.044 & 0.042 & 0.045 & 0.045 & 0.045 & 0.048 \\
        MMD-Miss       & 0.000 & 0.000 & 0.000 & 0.000 & 0.000 & 0.000 & 0.000 & 0.000 & 0.000 \\
        RISE           & 0.045 & 0.040 & 0.061 & 0.048 & 0.055 & 0.043 & 0.050 & 0.048 & 0.050 \\
        MMD            & 0.004 & 0.000 & 0.000 & 0.004 & 0.000 & 0.000 & 0.008 & 0.009 & 0.008 \\
        BD             & 0.062 & 0.047 & 0.050 & 0.058 & 0.054 & 0.047 & 0.044 & 0.048 & 0.043 \\
        MT             & 0.048 & 0.044 & 0.059 & 0.047 & 0.050 & 0.051 & 0.057 & 0.045 & 0.039 \\
        BRISE-c(sp)  & \textit{0.193} & \textit{0.200} & \textit{0.213} & \textit{0.281} & \textit{0.336} & \textit{0.391} & \textit{0.319} & \textit{0.375} & \textit{0.451} \\
        \hline
        $d$ & 10 & 20 & 50 & 10 & 20 & 50 & 10 & 20 & 50 \\
        \hline
        $ p_X = p_Y = 0.5$ & \multicolumn{3}{c|}{Setting I} & \multicolumn{3}{c|}{Setting II} & \multicolumn{3}{c|}{Setting III} \\
        \hline
        BRISE-c   & 0.062 & 0.048 & 0.057 & 0.045 & 0.043 & 0.056 & 0.040 & 0.052 & 0.057 \\   
        BRISE-v   & 0.062 & 0.061 & 0.049 & 0.068 & 0.064 & 0.055 & 0.067 & 0.055 & 0.048 \\   
        MMD-Miss  & 0.000 & 0.000 & 0.000 & 0.000 & 0.000 & 0.000 & 0.000 & 0.000 & 0.000 \\   
        RISE      & 0.040 & 0.055 & 0.056 & 0.044 & 0.049 & 0.048 & 0.048 & 0.044 & 0.055 \\   
        MMD       & 0.035 & 0.039 & 0.017 & 0.031 & 0.037 & 0.016 & 0.039 & 0.023 & 0.024 \\   
        BD        & 0.041 & 0.058 & 0.043 & 0.044 & 0.046 & 0.056 & 0.047 & 0.044 & 0.057 \\   
        MT        & 0.040 & 0.056 & 0.042 & 0.040 & 0.057 & 0.052 & 0.056 & 0.057 & 0.037 \\   
        MICE+RISE & \textit{0.906} & \textit{0.996} & \textit{1.000} & \textit{0.867} & \textit{0.981} & \textit{1.000} & \textit{0.914} & \textit{0.991} & \textit{1.000} \\   
        MICE+MMD  & \textit{0.530} & \textit{0.729} & \textit{0.915} & \textit{0.490} & \textit{0.627} & \textit{0.998} & \textit{0.596} & \textit{0.746} & \textit{0.967} \\   
        MICE+BD   & \textit{0.311} & \textit{0.381} & \textit{0.616} & \textit{0.217} & \textit{0.182} & \textit{0.392} & \textit{0.303} & \textit{0.233} & \textit{0.422} \\   
        MICE+MT   & 0.013 & \textit{0.177} & \textit{0.854} & 0.037 & 0.084 & \textit{0.433} & 0.014 & \textit{0.196} & \textit{0.886} \\   
        \hline
    \end{tabular}}
    \label{sim: Empirical sizes2}
\end{table}

\begin{table}[h!]
    \centering
    \caption{Estimated powers at significance level $\theta = 0.05$ with $m = n = 100$, $L = 2$, and $p_X = p_Y = 0.5$.
Results are shown for multivariate Gaussian (I), log-normal (II), and $t_5$ (III) distributions under location (a), scale (b), and mixed (c) alternatives. Top-1 and Top-2 powers are shown in \textbf{bold} and \underline{underlined}, respectively.}
    {
    \footnotesize
    \begin{tabular}{|c|ccc|ccc|ccc|}
        \hline
        $d$ & 200 & 500 & 1000 & 200 & 500 & 1000 & 200 & 500 & 1000 \\
        \hline
        Method & \multicolumn{3}{c|}{Setting I-a} & \multicolumn{3}{c|}{Setting I-b} & \multicolumn{3}{c|}{Setting I-c} \\
        \hline
        BRISE-c  & \textbf{0.859} & \textbf{0.737} & \textbf{0.583} & \textbf{0.849} & \textbf{0.922} & \textbf{0.957} & \textbf{0.961} & \textbf{0.943} & \textbf{0.934} \\
        BRISE-v  & \underline{0.659} & \underline{0.483} & \underline{0.357} & 0.537 & 0.675 & 0.762 & \underline{0.764} & \underline{0.730} & \underline{0.696} \\
        MMD-Miss & 0 & 0 & 0 & 0 & 0 & 0 & 0 & 0 & 0 \\
        RISE     & 0.326 & 0.242 & 0.201 & 0.464 & 0.599 & 0.695 & 0.300 & 0.268 & 0.279 \\
        MMD      & 0.111 & 0.003 & 0 & 0 & 0 & 0 & 0 & 0 & 0 \\
        BD       & 0.156 & 0.103 & 0.102 & \underline{0.624} & \underline{0.782} & \underline{0.877} & 0.083 & 0.070 & 0.094 \\
        MT       & 0.052 & 0.066 & 0.051 & 0.060 & 0.050 & 0.050 & 0.076 & 0.082 & 0.082 \\
        \hline
        Method & \multicolumn{3}{c|}{Setting II-a} & \multicolumn{3}{c|}{Setting II-b} & \multicolumn{3}{c|}{Setting II-c} \\
        \hline
        BRISE-c  & \underline{0.767} & \underline{0.564} & \underline{0.386} & \textbf{0.521} & \textbf{0.527} & \textbf{0.579} & \textbf{0.811} & \textbf{0.867} & \textbf{0.912} \\
        BRISE-v  & \textbf{0.817} & \textbf{0.680} & \textbf{0.473} & 0.282 & 0.262 & 0.262 & 0.617 & 0.630 & 0.639 \\
        MMD-Miss & 0 & 0 & 0 & 0 & 0 & 0 & 0 & 0 & 0 \\
        RISE     & 0.272 & 0.234 & 0.153 & 0.224 & 0.209 & 0.264 & 0.426 & 0.442 & 0.493 \\
        MMD      & 0.091 & 0.002 & 0 & 0.075 & 0.011 & 0.001 & 0.232 & 0.058 & 0.008 \\
        BD       & 0.074 & 0.065 & 0.063 & \underline{0.428} & \underline{0.478} & \underline{0.513} & \underline{0.642} & \underline{0.747} & \underline{0.800} \\
        MT       & 0.084 & 0.157 & 0.157 & 0.136 & 0.128 & 0.103 & 0.444 & 0.417 & 0.377 \\
        \hline
        Method & \multicolumn{3}{c|}{Setting III-a} & \multicolumn{3}{c|}{Setting III-b} & \multicolumn{3}{c|}{Setting III-c} \\
        \hline
        BRISE-c  & \underline{0.931} & \underline{0.834} & \underline{0.611} & \textbf{0.810} & \textbf{0.772} & \textbf{0.656} & \textbf{0.947} & \textbf{0.860} & \textbf{0.710} \\
        BRISE-v  & \textbf{0.976} & \textbf{0.938} & \textbf{0.777} & \underline{0.695} & \underline{0.648} & \underline{0.510} & \underline{0.854} & \underline{0.762} & \underline{0.620} \\
        MMD-Miss & 0 & 0 & 0 & 0 & 0 & 0 & 0 & 0 & 0 \\
        RISE     & 0.446 & 0.306 & 0.197 & 0.326 & 0.287 & 0.244 & 0.416 & 0.303 & 0.215 \\
        MMD      & 0.068 & 0.020 & 0.009 & 0.077 & 0.088 & 0.077 & 0.093 & 0.019 & 0.007 \\
        BD       & 0.054 & 0.041 & 0.041 & 0.352 & 0.378 & 0.386 & 0.058 & 0.042 & 0.043 \\
        MT       & 0.081 & 0.086 & 0.069 & 0.073 & 0.087 & 0.074 & 0.359 & 0.331 & 0.273 \\
        \hline
    \end{tabular}
    }
    \label{sim: power_default}
\end{table}

The estimated powers under the default setting are reported in Table~\ref{sim: power_default}. Across all dimensions and alternatives, the BRISE family consistently achieves higher power than the competing methods. In particular, BRISE-c shows the most stable performance, maintaining strong power across a wide range of distributional differences, while BRISE-v remains competitive, especially under heterogeneous alternatives.

\begin{figure}[ht]
    \centering
    \includegraphics[width=\linewidth]{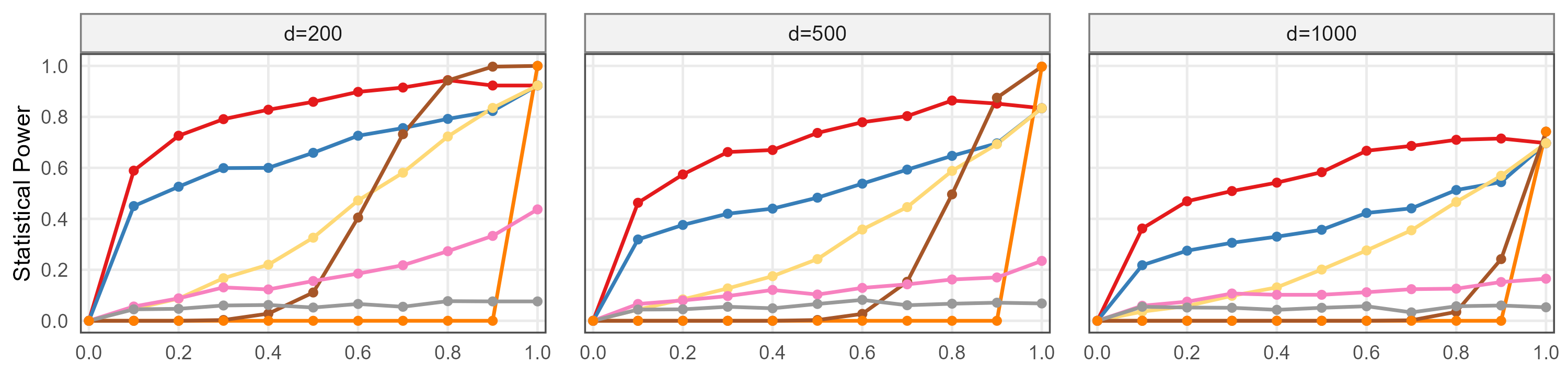}
    \includegraphics[width=\linewidth]{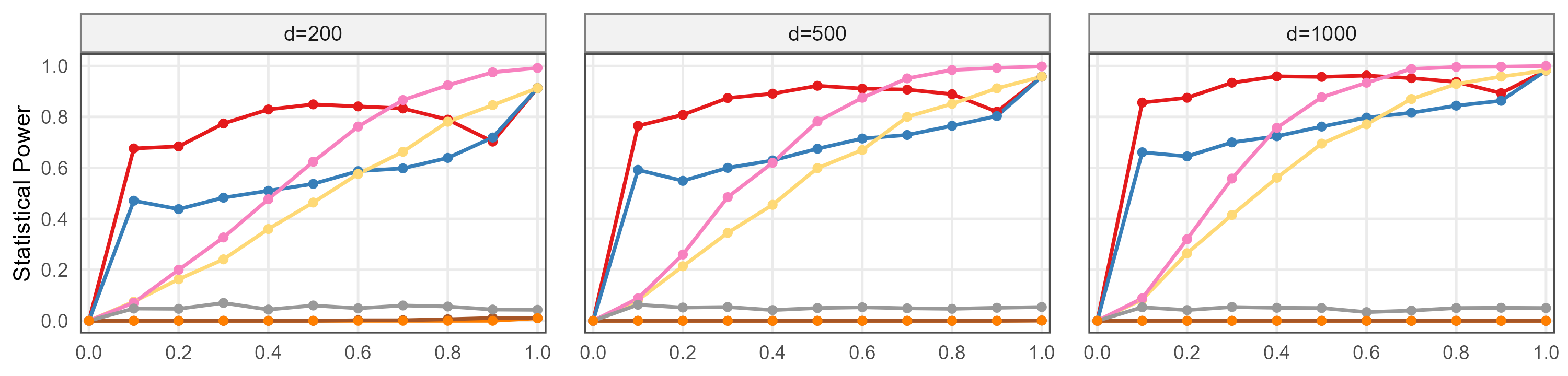}
    \includegraphics[width=\linewidth]{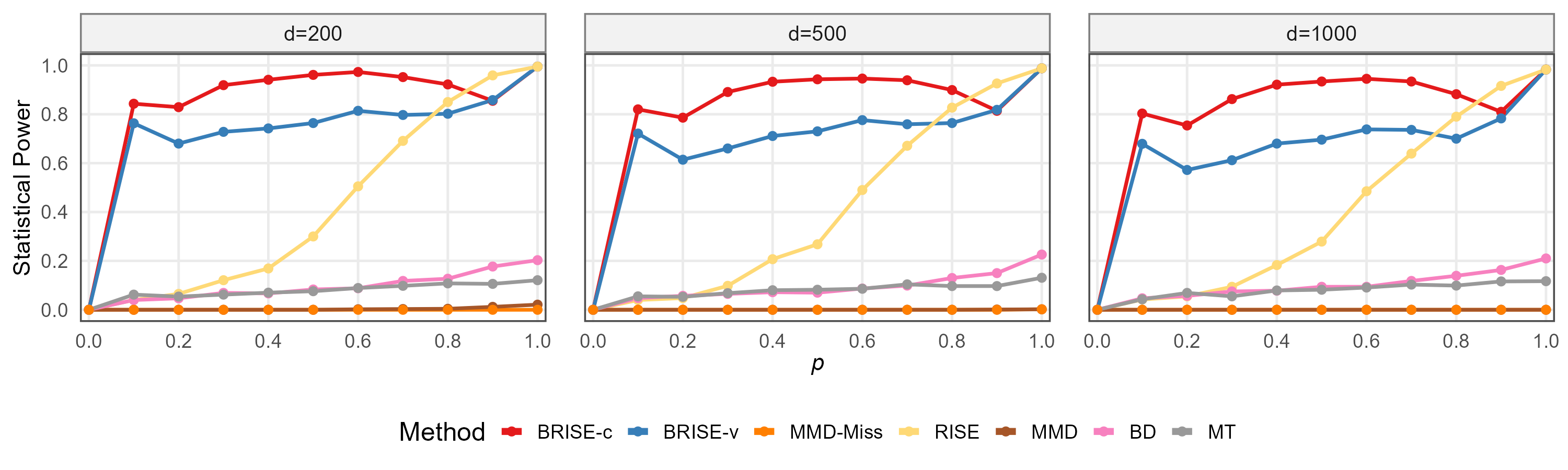}
    \caption{Trend of estimated powers vs. $p$ at significance level $\theta = 0.05$, for $m = n = 100$, $L = 2 $, $ p_X = p_Y = p$ and $d\in\{200,500,1000\}$. Top: Setting I-a; Middle: Setting I-b; Bottom: Setting I-c.}
    \label{fig:power-p}
\end{figure}

Figure~\ref{fig:power-p} further examines power as the sampling rates vary with $p_X = p_Y \in {0, 0.1, \ldots, 0.9, 1}$ under Setting~I. Both BRISE-c and BRISE-v are robust to decreasing sampling rates, with only modest power loss. In contrast, MMD-Miss, RISE, MMD, BD, and MT experience substantial power degradation or fail to perform reliably as sampling rates decrease.

When sampling rates are high, some baselines perform well for specific alternatives (MMD for location, BD for scale, and RISE for mixed alternatives). However, these methods rely on complete-case analysis and suffer from severe effective sample size reduction at low sampling rates, underscoring the robustness advantage of BRISE. MT and MMD-Miss show limited power across most settings.

Additional results under alternative distributions, imbalanced sample sizes, unequal sampling probabilities, and increased numbers of sources are provided in Supplementary~P. Across all scenarios, BRISE consistently maintains valid type-I error control and strong power, whereas complete-case methods often exhibit inflated or overly conservative type-I error or substantial power loss, provided Condition~\eqref{cond1} holds.

Overall, the simulation studies provide strong empirical support for the BPET framework and demonstrate that the BRISE tests are well suited for high-dimensional two-sample testing with block-wise missing data.

\section{Real Data Analysis}
\label{section: real data}
\subsection{Sepsis Identification Using Multi-source Biomarkers} 
\label{subsec: Spesis}

We illustrate the utility of BRISE using a sepsis study. Sepsis is a major public health concern associated with high morbidity and mortality \citep{singer2016third}. The goal of this analysis is to assess whether the distributions of age, sex, race, APACHE II score, absolute neutrophil count (ANC), absolute lymphocyte count (ALC), monocyte distribution width (MDW), and other biomarkers differ between septic patients and critically ill non-septic (CINS) patients.

Most existing studies of these biomarkers involve relatively small samples or focus on community-acquired sepsis in emergency department settings \citep{agnello2021monocyte}. In contrast, we consider hospital-acquired sepsis in surgical and trauma ICUs, where distinguishing sepsis from inflammation due to surgery or trauma is clinically challenging, and where patients are at elevated risk due to invasive procedures, indwelling devices, and prolonged hospitalization \citep{mas2021sepsis,cohen2023sepsis}.

The data were combined from three prospective observational studies of patients admitted to a surgical ICU at a quaternary-care academic medical center. Two studies were conducted by the UF Sepsis and Critical Illness Research Center (SCIRC) and registered at clinicaltrials.gov (NCT04414189, NCT05110937) \citep{brakenridge2021transcriptomic}, and the third study was part of an NIGMS-funded consortium at Shands-UF Hospital \citep{barrios2024adverse}. A key methodological challenge is block-wise missingness induced by study-specific protocols: biomarkers were measured on Days~1 and~4 in two studies, but on Day~2 in the third, and no study collected all biomarkers across all days. As a result, the combined dataset exhibits block-wise missing patterns that preclude standard complete-case or covariate-adjusted analyses.

\begin{figure}[htbp]
    \centering
    \includegraphics[width=\linewidth]{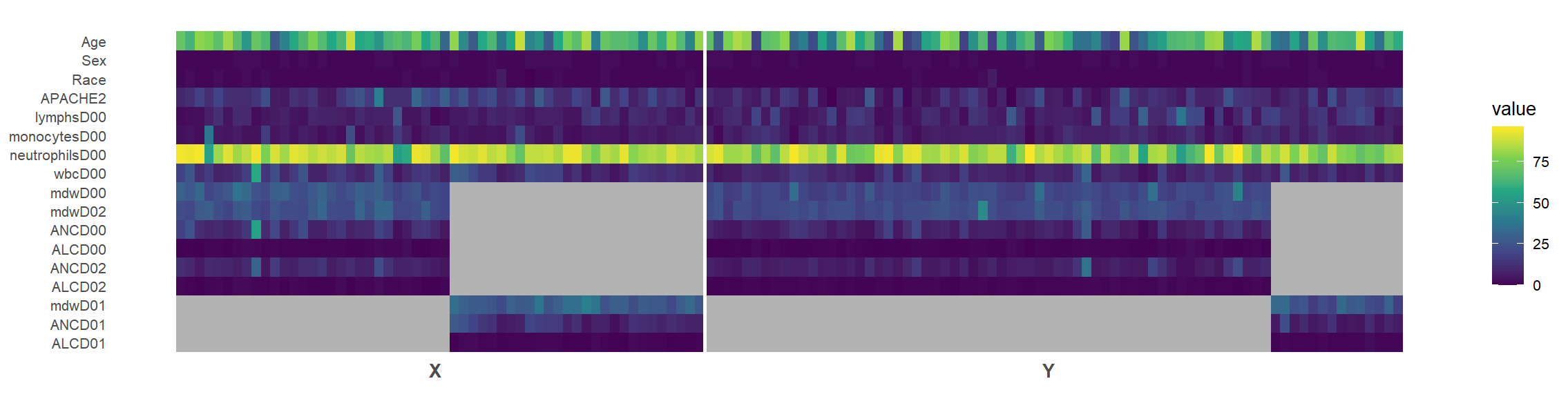}
    \caption{Heatmap of sepsis biomarker data. Gray indicates missing values.}
    \label{fig:heatmap of sepsis}
\end{figure}

The final dataset includes $56$ septic patients ($X$) and $74$ non-septic patients ($Y$). Figure~\ref{fig:heatmap of sepsis} provides a heatmap visualization of the biomarker data, with gray indicating missing values.

Table~\ref{tab:real data p-values} reports the $p$-values from BRISE and benchmark methods. Both BRISE-c and BRISE-v detect highly significant differences between septic and non-septic patients. In contrast, MMD-Miss fails to reject the null hypothesis, reflecting its conservative behavior in this setting, and complete-case analysis is infeasible due to the absence of fully observed samples. These results demonstrate the practical advantage of BRISE for real-world biomedical studies with block-wise missingness.

\begin{table}[htbp]
\caption{Reported $p$-values for the sepsis data; values $<0.05$ are \textbf{bolded}.}
\centering
\footnotesize
\begin{tabular}{|c|ccc|c|}
    \hline
    Method & BRISE-c & BRISE-v & MMD-Miss & CC Analysis\\
    \hline 
    Sepsis/CINS  & \textbf{5.38e-14} & \textbf{3.89e-15} & 1 & n.a. \\
    \hline
\end{tabular}
\label{tab:real data p-values}
\end{table}

\subsection{Multi-modal Differentiation of Alzheimer’s Disease}\label{subsec:ADNI realdata}

Distinguishing cognitive health states using heterogeneous medical data is a central application of multi-modal statistical analysis. In Alzheimer’s disease (AD), differences from cognitively normal (CN) subjects are often subtle and distributed across imaging, metabolic, and clinical biomarkers, making this problem well suited for evaluating methods designed for incomplete multi-source data.

We apply our methods to data from the Alzheimer’s Disease Neuroimaging Initiative (ADNI) \citep{ADNIa}. Each participant is identified by a unique Participant Roster ID (RID). Following \citet{balsis2015}, diagnostic groups are defined using Mini-Mental State Examination (MMSE) scores \citep{folstein1975mini}: subjects with average MMSE scores in $(24,30]$ are classified as CN, and those with scores $\le 24$ are classified as AD.

Three modalities are integrated in our analysis: Magnetic Resonance Imaging (MRI), Positron Emission Tomography (PET), and biospecimen data. MRI and PET features are preprocessed by the Center for Imaging of Neurodegenerative Diseases at the University of California, San Francisco, and the Jagust Lab at the University of California, Berkeley, respectively. Biospecimen data, generated by the Proteomics and Metabolomics Shared Resource at Duke University, include serum metabolite measurements such as carnitine and decanoylcarnitine. All data are accessed through the ADNI Analysis Ready Cohort Builder.\footnote{\url{https://ida.loni.usc.edu/explore/jsp/search_v2/search.jsp?project=ADNI}}

After merging modalities by RID, we remove cases with entrywise (random) missingness that do not conform to the modality-level block-wise structure assumed in this work. The resulting dataset consists of $2,555$ subjects with three modalities: MRI, PET, and biospecimen data (BIO), with dimensions
$d_{\mathrm{MRI}} = 289$,
$d_{\mathrm{PET}} = 318$, and
$d_{\mathrm{BIO}} = 518$. Approximately $89.3\%$ of subjects have at least one modality missing, and all seven non-empty modality combinations are observed; see Table~\ref{tab:adni-missing-patterns}.

\begin{table}[htbp]
\caption{Distribution of modality-level missingness patterns in the ADNI data.
Here $(S_{\mathrm{MRI}}, S_{\mathrm{PET}}, S_{\mathrm{BIO}})$ indicates whether each modality
is fully observed ($1$) or missing ($0$).}
\centering
\footnotesize
\begin{tabular}{|ccc|rr|}
    \hline
    $S_{\mathrm{MRI}}$ & $S_{\mathrm{PET}}$ & $S_{\mathrm{BIO}}$ & Count & Proportion \\
    \hline
    0 & 0 & 1 & 592 & 0.232 \\
    0 & 1 & 0 & 375 & 0.147 \\
    0 & 1 & 1 & 642 & 0.251 \\
    1 & 0 & 0 & 119 & 0.0466 \\
    1 & 0 & 1 & 17  & 0.00665 \\
    1 & 1 & 0 & 537 & 0.210 \\
    1 & 1 & 1 & 273 & 0.107 \\
    \hline
\end{tabular}
\label{tab:adni-missing-patterns}
\end{table}

We evaluate the methods on two datasets: (i) a balanced subset (Sub) with $200$ randomly selected subjects from each group (AD and CN), and (ii) the full dataset (Full) including all $2047$ CN and $507$ AD subjects. For each dataset, inference is conducted using permutation-based ($1000$ permutations) and, when available, asymptotic null distributions.

Table~\ref{tab:ADNI p-values} reports the resulting $p$-values. Both BRISE-c and BRISE-v detect highly significant differences between AD and CN subjects across all datasets and inference schemes. In contrast, RISE, MMD, and BD achieve significance only on the full dataset and fail to reject the null in the balanced subset. MMD-Miss yields $p$-values equal to $1$ in all settings, likely due to its inability to exploit cross-pattern information, while MT fails in both datasets. These results are consistent with our simulation findings and demonstrate the practical advantage of BRISE for high-dimensional, block-wise incomplete multi-modal data.

\begin{table}[htbp]
\caption{Reported $p$-values for the ADNI data; values $<0.05$ are \textbf{bolded}.}
\centering
\footnotesize
\begin{tabular}{|c|ccc|cccc|}
    \hline
    Method & BRISE-c & BRISE-v & MMD-Miss & RISE & MMD & BD & MT\\
    \hline 
    Sub(P) & \textbf{0} & \textbf{0} & 1 & 0.081 & 0.195 & 0.159 & 0.335\\
    Sub(A) & \textbf{3.56e-13} & \textbf{7.96e-11} & 1 & 0.074 & - & - & -\\
    Full(P) & \textbf{0} & \textbf{0} & 1 & \textbf{0.003} & \textbf{0.031} & \textbf{0.017} & 0.142\\
    Full(A) & \textbf{0} & \textbf{0} & 1 & \textbf{3.22e-05} & - & - & -\\
    \hline
\end{tabular}
\label{tab:ADNI p-values}
\end{table}

\section{Discussion and Conclusion}
\label{section: discussion}

In this work, we propose the general framework BPET for two-sample testing under block-wise missingness without requiring imputation or deletion of observations, accommodating possible MNAR. Further, we incorporate the rank-based RISE test within the proposed BPET framework to account for various types of modalities. We established the finite-sample properties, validity, and consistency of the proposed methods. Their effectiveness was demonstrated through both extensive simulations and real-world applications, and the BRISE exhibited strong empirical performance. Although developed for two-sample testing, BPET can be readily extended to other inferential tasks such as independence testing, clustering, and change-point detection in the presence of block-wise missingness.

As a general framework, we emphasize that BPET is not limited to graph/rank-based methods; a broad class of two-sample test procedures can be applied within the BPET framework. For example, by computing pattern-wise dissimilarities or kernel metrics (Section~\ref{subsec: BPET wf}), both kernel-based, and distance-based methods can be incorporated into BPET. 
In multi-modal data, samples from different patterns may have non-overlapping variables or modalities. In the main text, we proposed the zero-filling strategy in Section~\ref{subsec: BPET wf}; one may also adopt alternative measures, such as rank-based distances (Supplementary~M) or embedding all sources into a latent subspace (e.g., via partial multi-view clustering  following \citet[e.g.]{li2014partial,feng2024partial}) before computing distances on the aligned projections. An interesting observation from our simulation is that the proposed method leveraging the zero-filling strategy  consistently outperforms other dissimilarity measures. This raises a future research question: whether information from samples without shared or overlapping sources/modalities is important, and if so, how to efficiently leverage it. Our work is among the first to explore these important questions.

\bibliographystyle{chicago}
{\small \bibliography{Bibliography-MM-MC}}

\end{document}